\documentclass[journal,longauth]{aa}

\usepackage{graphicx}
\usepackage{amsmath}
\usepackage{amssymb}
\usepackage{verbatim}
\usepackage{array}
\usepackage{txfonts}
\usepackage{natbib}
\usepackage[switch]{lineno}
\usepackage{multirow}
\usepackage{float}
\usepackage{color}

\bibpunct{(}{)}{;}{a}{}{,}

\newcommand{\celltspace}{\rule{0pt}{2.8ex}}
\newcommand{\cellbspace}{\rule[-1.4ex]{0pt}{0pt}}

\def\deg{\ensuremath{^\circ}}

\newcommand{\msol}{\,M$_{\odot}$}

\newcommand{\wunit}{\,M$_{\odot}$\,yr$^{-1}$}

\newcommand{\gray}{$\gamma$-ray}
\newcommand{\grays}{$\gamma$-rays}

\defcitealias{Abdo:2010d}{Paper I}

\begin{document}

\title{Deep view of the Large Magellanic Cloud \\with six years of {\em Fermi}-LAT observations}
\titlerunning{A deep view of the Large Magellanic Cloud}
\author{\vspace{1mm}
M.~Ackermann$^{(1)}$ \and 
A.~Albert$^{(2)}$ \and 
W.~B.~Atwood$^{(3)}$ \and 
L.~Baldini$^{(4,2)}$ \and 
J.~Ballet$^{(5)}$ \and 
G.~Barbiellini$^{(6,7)}$ \and 
D.~Bastieri$^{(8,9)}$ \and 
R.~Bellazzini$^{(10)}$ \and 
E.~Bissaldi$^{(11)}$ \and 
E.~D.~Bloom$^{(2)}$
% ...
}
\authorrunning{LAT collaboration}

\institute{
\inst{1}~Deutsches Elektronen Synchrotron DESY, D-15738 Zeuthen, Germany\\ 
\inst{2}~W. W. Hansen Experimental Physics Laboratory, Kavli Institute for Particle Astrophysics and Cosmology, Department of Physics and SLAC National Accelerator Laboratory, Stanford University, Stanford, CA 94305, USA\\ 
\inst{3}~Santa Cruz Institute for Particle Physics, Department of Physics and Department of Astronomy and Astrophysics, University of California at Santa Cruz, Santa Cruz, CA 95064, USA\\ 
\inst{4}~Universit\`a di Pisa and Istituto Nazionale di Fisica Nucleare, Sezione di Pisa I-56127 Pisa, Italy\\ 
\inst{5}~Laboratoire AIM, CEA-IRFU/CNRS/Universit\'e Paris Diderot, Service d'Astrophysique, CEA Saclay, F-91191 Gif sur Yvette, France\\ 
\inst{6}~Istituto Nazionale di Fisica Nucleare, Sezione di Trieste, I-34127 Trieste, Italy\\ 
\inst{7}~Dipartimento di Fisica, Universit\`a di Trieste, I-34127 Trieste, Italy\\ 
\inst{8}~Istituto Nazionale di Fisica Nucleare, Sezione di Padova, I-35131 Padova, Italy\\ 
\inst{9}~Dipartimento di Fisica e Astronomia ``G. Galilei'', Universit\`a di Padova, I-35131 Padova, Italy\\ 
\inst{10}~Istituto Nazionale di Fisica Nucleare, Sezione di Pisa, I-56127 Pisa, Italy\\ 
\inst{11}~Istituto Nazionale di Fisica Nucleare, Sezione di Bari, I-70126 Bari, Italy\\ 
\inst{12}~Istituto Nazionale di Fisica Nucleare, Sezione di Torino, I-10125 Torino, Italy\\ 
\inst{13}~Dipartimento di Fisica Generale ``Amadeo Avogadro" , Universit\`a degli Studi di Torino, I-10125 Torino, Italy\\ 
\inst{14}~NASA Goddard Space Flight Center, Greenbelt, MD 20771, USA\\ 
\inst{15}~Laboratoire Univers et Particules de Montpellier, Universit\'e Montpellier, CNRS/IN2P3, Montpellier, France\\ 
\inst{16}~Laboratoire Leprince-Ringuet, \'Ecole polytechnique, CNRS/IN2P3, Palaiseau, France\\ 
\inst{17}~Consorzio Interuniversitario per la Fisica Spaziale (CIFS), I-10133 Torino, Italy\\ 
\inst{18}~INAF-Istituto di Astrofisica Spaziale e Fisica Cosmica, I-20133 Milano, Italy\\ 
\inst{19}~Agenzia Spaziale Italiana (ASI) Science Data Center, I-00133 Roma, Italy\\ 
\inst{20}~Istituto Nazionale di Fisica Nucleare, Sezione di Perugia, I-06123 Perugia, Italy\\ 
\inst{21}~Dipartimento di Fisica, Universit\`a degli Studi di Perugia, I-06123 Perugia, Italy\\ 
\inst{22}~College of Science, George Mason University, Fairfax, VA 22030, resident at Naval Research Laboratory, Washington, DC 20375, USA\\ 
\inst{23}~INAF Osservatorio Astronomico di Roma, I-00040 Monte Porzio Catone (Roma), Italy\\ 
\inst{24}~INAF Istituto di Radioastronomia, I-40129 Bologna, Italy\\ 
\inst{25}~Dipartimento di Astronomia, Universit\`a di Bologna, I-40127 Bologna, Italy\\ 
\inst{26}~Dipartimento di Fisica, Universit\`a di Udine and Istituto Nazionale di Fisica Nucleare, Sezione di Trieste, Gruppo Collegato di Udine, I-33100 Udine\\ 
\inst{27}~Universit\`a Telematica Pegaso, Piazza Trieste e Trento, 48, I-80132 Napoli, Italy\\ 
\inst{28}~Universit\`a di Udine, I-33100 Udine, Italy\\ 
\inst{29}~Dipartimento di Fisica ``M. Merlin" dell'Universit\`a e del Politecnico di Bari, I-70126 Bari, Italy\\ 
\inst{30}~Centre d'\'Etudes Nucl\'eaires de Bordeaux Gradignan, IN2P3/CNRS, Universit\'e Bordeaux 1, BP120, F-33175 Gradignan Cedex, France\\ 
\inst{31}~Laboratoire de Physique et Chimie de l'Environnement et de l'Espace -- Universit\'e d'Orl\'eans / CNRS, F-45071 Orl\'eans Cedex 02, France\\ 
\inst{32}~Station de radioastronomie de Nan\c{c}ay, Observatoire de Paris, CNRS/INSU, F-18330 Nan\c{c}ay, France\\ 
\inst{33}~NASA Postdoctoral Program Fellow, USA\\ 
\inst{34}~School of Physics and Astronomy, University of Southampton, Highfield, Southampton, SO17 1BJ, UK\\ 
\inst{35}~Science Institute, University of Iceland, IS-107 Reykjavik, Iceland\\ 
\inst{36}~CNRS, IRAP, F-31028 Toulouse cedex 4, France\\ 
\inst{37}~Universit\'e de Toulouse, UPS-OMP, IRAP, Toulouse, France\\ 
\inst{38}~Department of Physics, KTH Royal Institute of Technology, AlbaNova, SE-106 91 Stockholm, Sweden\\ 
\inst{39}~The Oskar Klein Centre for Cosmoparticle Physics, AlbaNova, SE-106 91 Stockholm, Sweden\\ 
\inst{40}~Institute of Space Sciences (IEEC-CSIC), Campus UAB, E-08193 Barcelona, Spain\\ 
\inst{41}~Hiroshima Astrophysical Science Center, Hiroshima University, Higashi-Hiroshima, Hiroshima 739-8526, Japan\\ 
\inst{42}~Istituto Nazionale di Fisica Nucleare, Sezione di Roma ``Tor Vergata", I-00133 Roma, Italy\\ 
\inst{43}~Center for Cosmology, Physics and Astronomy Department, University of California, Irvine, CA 92697-2575, USA\\ 
\inst{44}~Department of Physics and Astronomy, University of Denver, Denver, CO 80208, USA\\ 
\inst{45}~Max-Planck-Institut f\"ur Physik, D-80805 M\"unchen, Germany\\ 
\inst{46}~Funded by contract FIRB-2012-RBFR12PM1F from the Italian Ministry of Education, University and Research (MIUR)\\ 
\inst{47}~Institut f\"ur Astro- und Teilchenphysik and Institut f\"ur Theoretische Physik, Leopold-Franzens-Universit\"at Innsbruck, A-6020 Innsbruck, Austria\\ 
\inst{48}~Department of Physics, Stockholm University, AlbaNova, SE-106 91 Stockholm, Sweden\\ 
\inst{49}~NYCB Real-Time Computing Inc., Lattingtown, NY 11560-1025, USA\\ 
\inst{50}~Department of Chemistry and Physics, Purdue University Calumet, Hammond, IN 46323-2094, USA\\ 
\inst{51}~Department of Physical Sciences, Hiroshima University, Higashi-Hiroshima, Hiroshima 739-8526, Japan\\ 
\inst{52}~Instituci\'o Catalana de Recerca i Estudis Avan\c{c}ats (ICREA), Barcelona, Spain\\ 
\inst{53}~Department of Physics and Department of Astronomy, University of Maryland, College Park, MD 20742, USA\\ 
\email{pierrick.martin@irap.omp.eu} \\
}
\date{Received: 09 July 2015 / Accepted: 28 September 2015}
\abstract{\mbox{The nearby Large Magellanic Cloud (LMC) provides a rare opportunity of a spatially resolved view} of an external star-forming galaxy in \grays. The LMC was detected at 0.1--100\,GeV as an extended source with CGRO/EGRET and using early observations with the {\em Fermi}-LAT. The emission was found to correlate with massive star-forming regions and to be particularly bright towards 30 Doradus.}
{Studies of the origin and transport of cosmic rays (CRs) in the Milky Way are frequently hampered by line-of-sight confusion and poor distance determination. The LMC offers a complementary way to address these questions by revealing whether and how the \gray\ emission is connected to specific objects, populations of objects, and structures in the galaxy.}
{\mbox{We revisited the \gray\ emission from the LMC using about 73 months of {\em Fermi}-LAT P7REP data in the 0.2--100\,GeV range.} We developed a complete spatial and spectral model of the LMC emission, for which we tested several approaches: a simple geometrical description, template-fitting, and a physically driven model for CR-induced interstellar emission.}
{In addition to identifying PSR~J0540$-$6919 through its pulsations,
we find two hard sources positionally coincident with plerion N~157B and supernova remnant \object{N~132D}, which were also detected at TeV energies with H.E.S.S. We detect an additional soft source that is currently unidentified. Extended emission dominates the total flux from the LMC. It consists of an extended component of about the size of the galaxy and additional emission from three to four regions with degree-scale sizes. If it is interpreted as CRs interacting with interstellar gas, the large-scale emission implies a large-scale population of $\sim$1--100\,GeV CRs with a density of $\sim$30\% of the local Galactic value. On top of that, the three to four small-scale emission regions would correspond to enhancements of the CR density by factors 2 to 6 or higher, possibly more energetic and younger populations of CRs compared to the large-scale population. An alternative explanation is that this is emission from an unresolved population of at least two dozen objects, such as pulsars and their nebulae or supernova remnants. This small-scale extended emission has a spatial distribution that does not clearly correlate with known components of the LMC, except for a possible relation to cavities and supergiant shells.}
{The {\em Fermi}-LAT GeV observations allowed us to detect individual sources in the LMC. Three of the newly discovered sources are associated with rare and extreme objects. The 30 Doradus region is prominent in GeV \grays\ because PSR~J0540$-$6919 and N~157B are strong emitters. The extended emission from the galaxy has an unexpected spatial distribution, and observations at higher energies and in radio may help to clarify its origin.}

\maketitle

\section{Introduction}
\label{intro}

The \object{Large Magellanic Cloud} (LMC) is a unique target for high-energy astrophysics. Given the sensitivity and angular resolution of current \gray\ instruments, it is the best opportunity we have to obtain an external and spatially resolved view of a star-forming galaxy. The \object{Small Magellanic Cloud} (SMC) offers a similar opportunity, but the galaxy is smaller, farther away, less active in terms of star formation, and has a more complex and less favourable geometry \citep{Scowcroft:2015}. Other star-forming galaxies within reach of current \gray\ instruments are at least an order of magnitude more distant \citep[e.g. M31, M82, and NGC 253;][]{Abdo:2010f,Abdo:2009l,Abramowski:2012} and have not been spatially resolved (in either the GeV or
TeV domain). 

In the Milky Way, hundreds of discrete objects of different classes have been detected, together with diffuse interstellar emission over a broad range of spatial scales. Studies of Galactic sources are often complicated by line-of-sight confusion and inaccurate distance estimates, however. In contrast, the LMC offers more modest prospects in terms of the number of detectable sources, but it provides an outside and global perspective on a whole galaxy seen at low inclination, whose distance is determined at the percent level \citep{Pietrzynski:2013}. 

Studying the LMC in \grays\ can be a valuable complement to Milky Way studies on a variety of topics: particle accelerators such as supernova remnants (SNRs), pulsars,  and pulsar wind nebulae (PWNe), with the possibility to find more rare and extreme objects; diffuse interstellar emission, which may show us how a cosmic-ray population developed in another galaxy and thus may be a useful test of our current theories on cosmic-ray transport; indirect searches for dark matter, which provides limits competitive with recent constraints, and that have different uncertainties and systematics \citep{Buckley:2015}. On the first two subjects, the outside point of view may reveal how cosmic-ray sources and transport are related to the structure and content of a galaxy as a whole. Here, building a global picture will be facilitated by the fact that the LMC was deeply surveyed over the past decades by almost all instruments that were able to do so, from radio to \grays.

From 11 months of {\em Fermi}-LAT continuous all sky-survey observations, \citet{Abdo:2010d}, hereafter Paper I, reported the 33$\sigma$ detection of the LMC in \grays. The source was spatially extended and its spectrum was consistent with the observed \grays\ originating from cosmic rays (CRs) interacting with the interstellar medium (ISM) through inverse-Compton scattering, Bremsstrahlung, and hadronic interactions, but contributions from discrete objects such as pulsars were not ruled out at that time. The emission was found to be relatively strong in the direction of the \object{30 Doradus} star-forming region; more generally, the emission seemed spatially correlated with tracers of massive star-forming regions, such as the H$\alpha$ emission or the population of known Wolf-Rayet stars, and showed little correlation with the gas column density distribution. This was interpreted as evidence supporting the idea that CRs are accelerated in massive star-forming regions as a result of the large amounts of kinetic energy released by the stellar winds and supernova explosions of massive stars. In addition, the close confinement of \gray\ emission to star-forming regions suggested a relatively short GeV CR proton diffusion length. 

We now have over six times more data than were used in \citetalias{Abdo:2010d} and better instrument performance. In this paper, we present a new analysis of the GeV \gray\ emission from the LMC. The paper is organised as follows: we first introduce the data set used in this work (Sect. \ref{data}) and the models developed to account for the emission seen in the LMC region (Sect. \ref{model}); then, the properties of the point-like sources found within the LMC boundaries are introduced and their identification or association with known objects is discussed (Sect. \ref{res_ptsrc}); last, the extended \gray\ emission from the LMC is presented, and a possible interpretation in terms of CR-induced interstellar emission or unresolved population of sources is given (Sect. \ref{res_extsrc}). Throughout the paper, the distance of the LMC is assumed to be $d=50.1$\,kpc \citep{Pietrzynski:2013}, and its inclination with respect to the line of sight is assumed to be $i=30\deg$ \citep{vanderMarel:2006}.

\newpage

\section{Data selection}
\label{data}

The following analysis was performed using 73.3 months of observations with the {\em Fermi} LAT (mission elapsed time 239587200 to 432694964), primarily taken in all-sky survey mode \citep{Atwood:2009}. The data set was produced with the so-called Pass 7 reprocessed (P7REP) version of the event analysis and selection criteria, which takes into account effects measured in flight that were not considered in pre-launch performance estimates, such as pile-up and accidental coincidence effects in the detector subsystems \citep{Ackermann:2012c}, and updated calibration constants, to include effects such as the degradation in the calorimeter light yield \citep{Bregeon:2013}. SOURCE class events were used in this work, excluding those coming from zenith angles larger than 100\deg\ or detected when the rocking angle of the satellite was larger than 52\deg\  to reduce contamination by atmospheric \grays\ from the Earth. We analysed together events converting in the front and back sections of the LAT, but we checked that considering them separately does not alter the results. The analysis was performed using Science Tools package v09r32p05 and using the P7REP\_SOURCE\_V15 instrument response functions (IRFs), all available from the {\em Fermi} Science Support Center\footnote{See http://fermi.gsfc.nasa.gov/ssc/}.

\begin{figure}[!t]
\begin{center}
\includegraphics[width= 8cm]{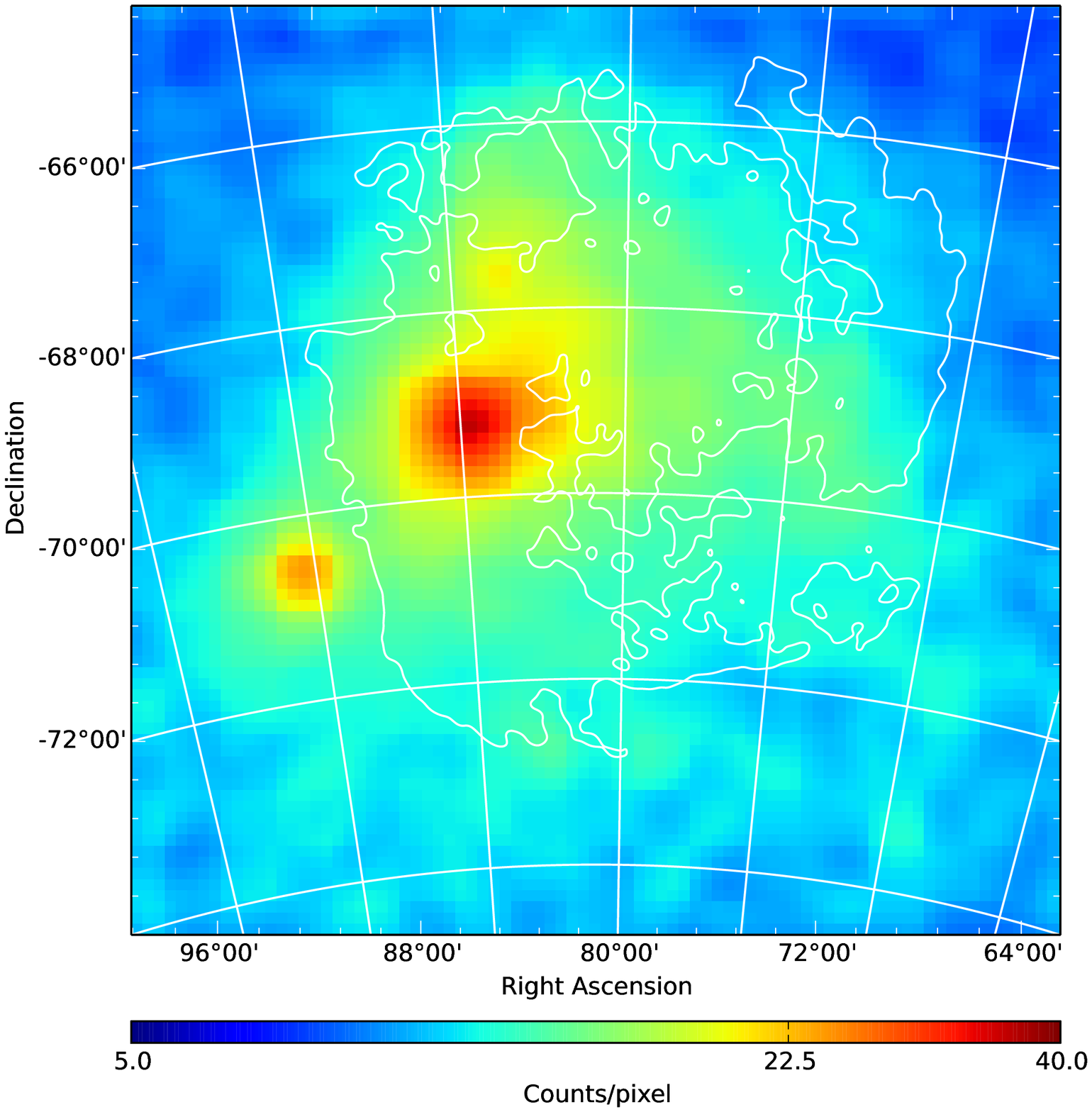}
\includegraphics[width= 8cm]{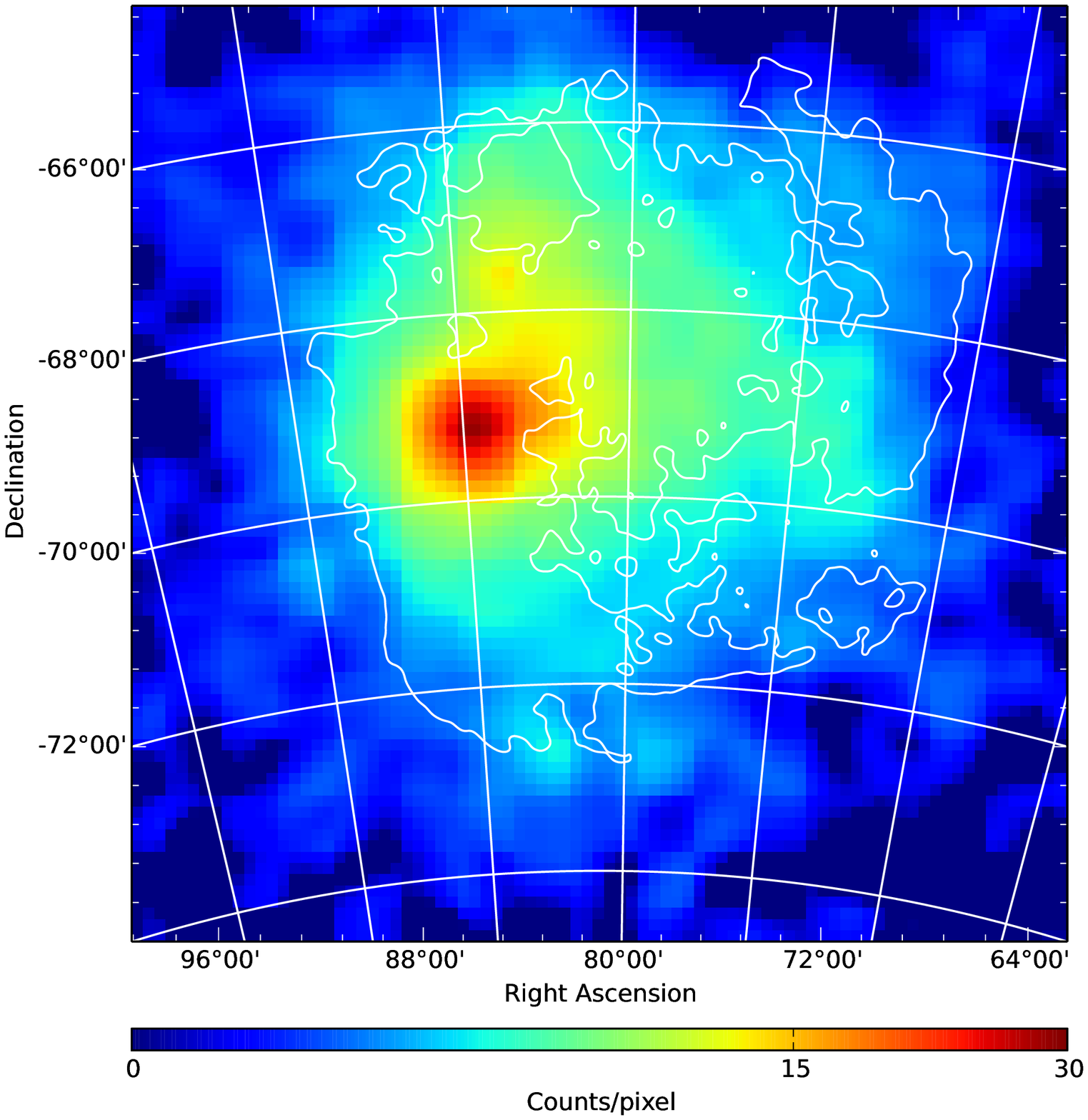}
\caption{Total counts map in the 0.2--100\,GeV band and residual counts map after subtracting the background model described in Sect. \ref{model_bgd} (top and bottom panels, respectively). Both maps have $0.1\deg \times 0.1\deg$ pixels and were smoothed with a Gaussian kernel with $\sigma=0.2\deg$. Colours are displayed on a square-root scale. White lines are contours of the atomic hydrogen distribution in the LMC at a relative value of 1/8 of the peak in the distribution (see Sect. \ref{model_emiss} for the origin of the data).}
\label{fig_cntmap}
\end{center}
\end{figure}

A region of interest (ROI) specific to the LMC was defined as a $10\deg \times 10\deg$ square centred on $(\alpha,\delta)=(80.894\deg,-69.756\deg)$ and aligned on equatorial coordinates (here and throughout the paper, equatorial coordinates correspond to the J2000.0 epoch)\footnote{As reported below, using a larger ROI had no impact on the results.}. Events contained in this region were selected from the data set described above. The energy range considered in this analysis is 0.2--100\,GeV; the lower energy bound is dictated by the poor angular resolution at the lowest energies. A counts map of the ROI is shown in the top panel of Fig. \ref{fig_cntmap}.

\section{Emission modelling}
\label{model}

We aim at building a complete model for the spatial and spectral distribution of the counts in the ROI and energy range defined above. As summarised below, we tested several possible approaches, each providing different insights into the \gray\ emission properties. All rely on a model-fitting procedure following a maximum likelihood approach for binned data and Poisson statistics. A given model consists of several emission components and has a certain number of free parameters. A distribution of expected counts in position and energy is obtained by convolving the model with the IRFs, taking into account the exposure achieved for the data set that is used. The free parameters are then adjusted in an iterative way until the distribution of expected counts provides the highest likelihood of the data given the model. Fits were made over the 0.2--100\,GeV range with 17 logarithmic energy bins and using $0.1\deg \times 0.1\deg$ pixels. In the following, models are compared based on the maximum value of the logarithm of the likelihood function, denoted $\log \mathcal L$. The significance of model components or additional parameters is evaluated using the test statistic, whose expression is $TS = 2 (\log \mathcal L - \log \mathcal L_0)$, where $L_0$ is the likelihood of the reference model without the additional parameter or component. A summary of the likelihoods and numbers of degrees of freedom of each model is given in Sect. \ref{model_sum}.

\newpage

\subsection{Background model}
\label{model_bgd}

As a first step in the process of modelling the emission over the ROI and before developing a model for the LMC, we have to account for known background and foreground emission in the form of diffuse or isolated sources. These are:

1- The Galactic interstellar emission, arising from CRs interacting with the ISM in our Galaxy. In the {\em Fermi}-LAT energy range, this emission is dominated by hadronic emission from interstellar gas \citep{Ackermann:2012b}. Even at the Galactic latitude of the LMC, $b \sim -33\deg$, this foreground radiation is clearly present as structured emission in the counts map. We modelled it using the template provided by the {\em Fermi} Science Support Center, \textit{gll\_iem\_v05\_rev1.fit}, with free normalisation in the fit. In the preparation of this template \citep{Casandjian:2015}, any signal from gas in the LMC was removed, so that the \gray\ emission from the LMC is not erroneously absorbed in the Galactic diffuse emission model. Importantly, the LMC region is not affected by the large-scale residual structures re-injected into the template.

2- An isotropic background, which accounts for an approximately isotropic diffuse \gray\ emission component and residual CRs misclassified as \grays\ in the LAT. The origin of the astrophysical emission is currently unclear and it may come from multiple sources, ranging from the solar system to cosmological structures \citep{Ackermann:2015}. It was modelled using the publicly available isotropic spectral model \textit{iso\_source\_v05.txt}, with free normalisation in the fit.

3- All isolated sources in the region that were previously detected and listed in the {\em Fermi}-LAT second source catalogue \citep[2FGL, ][]{Nolan:2012}. A total of seven sources fall into the ROI defined above. Four of them were dismissed because they are located within the LMC boundaries and may actually correspond to components of the LMC emission that we aimed at modelling. Source 2FGL~J0532.5$-$7223 was excluded because its significance was too low (below 3$\sigma$), but a source not listed in the 2FGL catalogue was found nearby, at the position $(\alpha,\delta)=(82.4^\circ,-72.7^\circ)$, with a $TS$ above 100, a power-law spectrum, and a variable flux. This additional source is present in the {\em Fermi}-LAT third source catalogue as 3FGL~J0529.8$-$7242 \citep{Acero:2015}. The other field sources are 2FGL~J0438.0$-$7331 and 2FGL~J0601.1$-$7037, the latter being associated with the radio source PKS~0601$-$70 and exhibiting strongly variable emission. All three were included in the model as point-like sources using the spectral shapes identified as most suitable in the 2FGL catalogue and leaving their spectral parameters free in the fit. We also included sources lying outside the ROI, up to a distance of 3\deg,  to account for spillover of their emission inside the ROI at low energies, where the point-spread function has a degree-scale size. A total of ten such sources were included in the model, with spectral shapes and parameters fixed at the catalogue values.

All the components described above form the basis of the emission model and are referred to as the background model. It has a total of nine degrees of freedom (one for each diffuse emission template, three for 2FGL~J0601.1$-$7037, and two each for 2FGL~J0438.0$-$7331 and 3FGL~J0529.8$-$7242). We now describe the modelling of the excess signal that is not accounted for by this background model.

\begin{table*}[!t]
\begin{minipage}[][4cm][c]{\textwidth}
\begin{center}
\caption{Point sources found in the LMC}
\label{tab_ptsrc}
\begin{tabular}{|c|c|c|c|c|}
\hline
\celltspace Name & Coordinates & Error radius & Spectrum & $TS$ \cellbspace \\
\hline
 P1 & 85.0465\deg, -69.3316\deg &  & PLC & 151  \\
\hline
 P2 & 84.43\deg, -69.17\deg & 0.02\deg & PL & 96  \\
\hline
 P3 & 83.54\deg, -67.54\deg & 0.06\deg & PL & 116  \\
\hline
 P4 & 81.13\deg, -69.62\deg & 0.03\deg & PL & 28  \\
\hline
\end{tabular}
\end{center}
From left to right, the columns list the name of the point source (P1 being firmly identified as PSR~J0540$-$6919), equatorial coordinates, error on the position, best-fit spectral shape (PL=power law, PLC=power law with exponential cutoff), and the test statistic of the source. All values correspond to the emissivity model. Position uncertainties correspond to a 95\% confidence level.
\end{minipage}
\end{table*}

\begin{table*}[!t]
\begin{minipage}[][4.5cm][c]{\textwidth}
\begin{center}
\caption{Emission components of the analytic model}
\label{tab_anamodel}
\begin{tabular}{|c|c|c|c|c|c|}
\hline
\celltspace Name & Coordinates & Error radius & Size & Spectrum & $TS$ \cellbspace \\
\hline
 G1 & 79.70\deg, -68.55\deg & 0.2\deg & $1.95\deg \pm 0.05\deg$ & LP & 905  \\
\hline
 G2 & 83.40\deg, -69.15\deg & 0.1\deg & $0.70\deg \pm 0.05\deg$  & PL & 251  \\
\hline
 G3 & 83.10\deg, -66.60\deg & 0.2\deg & $0.45\deg \pm 0.10\deg$  & PL & 211 \\
\hline
 G4 & 74.20\deg, -69.50\deg & 0.3\deg & $0.40\deg \pm 0.10\deg$  & PL & 55  \\
\hline
\end{tabular}
\end{center}
From left to right, the columns list the name of the 2D Gaussian intensity distribution component, equatorial coordinates of the centre of the 2D Gaussian, error on the position, $\sigma$ parameter of the 2D Gaussian, best-fit spectral shape (PL=power law, LP=log-parabola),
and the test statistic of the emission component. Position uncertainties correspond to a 95\% confidence level, size uncertainties correspond to 1$\sigma$.
\end{minipage}
\end{table*}

\begin{table*}[!t]
\begin{minipage}[][5cm][c]{\textwidth}
\begin{center}
\caption{Emissivity components of the emissivity model}
\label{tab_emimodel}
\begin{tabular}{|c|c|c|c|c|c|}
\hline
\celltspace Name & Coordinates & Error radius & Size & Spectrum & $TS$ \cellbspace \\
\hline
 E0 & 80.00\deg, -68.00\deg & 0.5\deg & $3.0\deg$ & LP & 891   \\
\hline
 E1 & 82.40\deg, -68.85\deg & 0.4\deg & $0.6\deg \pm 0.1\deg$  & PL & 276   \\
\hline
 E2 & 82.97\deg, -66.65\deg & 0.2\deg & $0.4\deg \pm 0.1\deg$  & PL & 270   \\
\hline
 E3 & 82.25\deg, -69.25\deg & 0.2\deg & $0.3\deg \pm 0.1\deg$  & PL & 276   \\
\hline
 E4 & 75.25\deg, -69.75\deg & 0.4\deg & $0.6\deg \pm 0.2\deg$  & PL & 120   \\
\hline
\end{tabular}
\end{center}
From left to right, the columns list the name of the emissivity component, equatorial coordinates of the centre of the 2D Gaussian emissivity distribution, error on the position, $\sigma$ parameter of the 2D Gaussian emissivity distribution, best-fit spectral shape (PL=power law, LP=log-parabola), and the test statistic of the emission component (the $TS$ for E1 and E3 is that obtained when both components are tied together, see text). Position uncertainties correspond to a 95\% confidence level, size uncertainties correspond to 1$\sigma$.
\end{minipage}
\end{table*}

\subsection{Analytic model}
\label{model_ana}

Starting from the background model, we first aim to describe the remaining emission with a combination of point-like and 2D Gaussian-shaped spatial intensity distributions, adding new components successively. 

Point-like sources can be identified if they have hard spectra and are bright enough, because the angular resolution at high energies $>$10--20\,GeV is relatively good and allows distinguishing them from any extended emission. Inspection of the $>$20\,GeV counts map suggested the presence of two such sources (called P2 and P4 in the following), the significance and point-like nature of which was confirmed by subsequent analyses. In addition, one source was identified as a \gray\ pulsar from its characteristic pulsations: the source called P1 in the following was identified as \object{PSR~J0540$-$6919} (see Sect. \ref{res_pt_psr0540}), and its position was fixed to the position of the pulsar from optical observations \citep{Mignani:2010}. Starting from these three point-like contributions to the LMC emission, an iterative procedure was used to characterise the remaining emission.

At each step, a scan over position and size of the new emission component is performed to identify the source that provides the best fit to the data. 
For each trial position and size, the full model is multiplied by the exposure, convolved with the point spread function, and fit to the data in a binned maximum likelihood analysis for Poisson statistics. A power-law spectral shape is assumed for the new component (which turned out to be a good approximation for most components). If the improvement of the likelihood is significant -- which we defined as $TS \geq 25$ -- the component is added to the model and a new iteration starts. The process stops when adding a new component yields a $TS < 25$. The subsequent step is to re-optimize the positions and sizes of the components, from the brightest to the faintest in turn. The final stage is deriving bin-by-bin spectra for all components to check that the initially adopted power-law spectral shape is appropriate. If not, it is replaced by a power-law with exponential cutoff or a log-parabola shape, depending on which provides the best fit and a significant improvement over the power-law assumption \citep[for the formulae of the different spectral functions, see][]{Nolan:2012}.

This procedure resulted in an emission model with eight components: four point-like objects and four Gaussian-shaped, spatially extended components. In the following, this model is referred to as the analytic model.
The properties of all components are summarised in Tables \ref{tab_ptsrc} and \ref{tab_anamodel}, and their layout is illustrated in Fig. \ref{fig_charts} (the positions listed in Table \ref{tab_ptsrc} correspond to the emissivity model described below; the positions corresponding to the analytic model are very similar and fully consistent with them, as illustrated in Fig. \ref{fig_loca_ptsrc}). Point sources are labelled P1 to P4, and extended sources are labelled G1 to G4.

\subsection{Emissivity model}
\label{model_emiss}

The diffuse emission part of the analytic model includes a large-scale contribution, which is located close to the geometrical centre of the LMC disk and covers a large part of its area, and three well-separated smaller-size components. It is reasonable to assume that the extended sources correspond to populations of CRs interacting with the ISM, and we explored this possibility with a dedicated modelling.

Under this hypothesis, the emission model was refined by adopting a more physical approach to determine the extended emission components. In the $\sim$0.1--100\,GeV range, the interstellar radiation is dominated by gas-related processes, hadronic interactions especially \citep[see for instance][]{Ackermann:2012}. The intensity in a given direction can be expressed as the product of an emissivity $q(E_{\gamma})$, the \gray\ emission rate per hydrogen atom per unit energy per solid angle (which depends on CR density and spectrum and on nuclear interaction physics), and a hydrogen gas column density (assuming that all gas is pervaded by the same CR flux). Instead of searching for a best-fit combination of 2D Gaussian-shaped intensity distributions, we therefore performed an iterative search for a combination of 2D Gaussian-shaped emissivity components, which, when multiplied with the gas column density distribution of the LMC, provides the best fit to the data. We refer to this second model as the emissivity model. The gas column density map used in this work includes atomic, molecular, and ionized hydrogen, and its preparation is described in \citetalias{Abdo:2010d}. For the dominant atomic and molecular hydrogen species, the map is based on ATCA plus Parkes and NANTEN data, respectively, assuming optically thin emission for atomic hydrogen, and a CO intensity to H$_2$ column density conversion factor of $X_{\textrm{CO}} = 7 \times 10^{20}$\,H$_2$\,cm$^{-2}$/(K\,km\,s$^{-1}$).

\begin{table}[!t]
\begin{center}
\caption{Comparison of the emission models}
\label{tab_allmodels}
\begin{tabular}{|c|c|c|c|c|c|}
\hline
\celltspace Model & $\log \mathcal L$ & $TS$ & N$_{\textrm{dof}}$  \cellbspace \\
\hline
Background & -65219.9 & - & 9   \\
\hline
Analytic & -59167.7 & 12104.4 & 45   \\
\hline
Emissivity & -59131.1 & 12177.6 & 47   \\
\hline
Ionised gas + PSR~J0540$-$6919 & -59462.1 & 11515.6 & 15   \\
\hline
Ionised gas + 4 point sources & -59400.0 & 11639.8 & 27   \\
\hline
All gas + PSR~J0540$-$6919 & -60062.3 & 10315.2 & 15   \\
\hline
All gas + 4 point sources & -59548.9 & 11342.0 & 27   \\
\hline
\end{tabular}
\end{center}
From left to right, the columns list the name of the model, log(likelihood) values obtained in the model fitting (not including the sum of counts term that is model-independent), corresponding test statistic with respect to the background model, and the number of degrees of freedom in the model.
\end{table}

In this context, the large-scale component G1 of the analytic model is interpreted as arising from a large-scale population of CRs spread almost uniformly across the LMC disk and interacting with the gas. An assumption in building the emissivity model is that there is a large-scale CR population in the LMC, which is described with a 2D Gaussian emissivity distribution initially centred on $(\alpha,\delta)=(80.0^\circ,-68.5^\circ)$ and having a fixed width $\sigma=3.0^\circ$ (approximately the angular radius of the galaxy; using a flat emissivity profile over the entire extent of the LMC was found to degrade the fit by 38 in $\log \mathcal L$). We then searched for additional emissivity components on top of that. The method is very similar to the technique used to build the analytic model: at each iteration, a scan over position and size is performed to identify a new emissivity component that provides the best fit to the data and a significant improvement over the previous iteration. In the fits, a log-parabola spectral shape is assumed for the new component because it mimics a local interstellar emission spectrum in the $\sim$0.1--10\,GeV range fairly well. After all components were added, we re-optimised the positions and sizes of the components, from the brightest to the faintest in turn, including point-like sources (except P1). We also optimised the position of the large-scale CR component and found a best-fit position close to our initial choice. We then revised the log-parabola spectral shape assumption. All components could be described by a simple power law (see Sect. \ref{res_ext_spec}), except for the large-scale component, which was found to be more satisfactorily described by a log-parabola or a tabulated function derived from the local gas emissivity spectrum (see Sect. \ref{res_ext_spec}).

\begin{table*}[!ht]
\begin{minipage}[][5.2cm][c]{\textwidth}
\begin{center}
\caption{Best-fit spectral parameters of the point-like sources}
\label{tab_ptsrcspec}
\begin{tabular}{|c|c|c|c|}
\hline
\celltspace Source & Parameters & F$_{200}$ & F$_{100}$ \cellbspace \\
\hline
\celltspace P1 (PSR~J0540$-$6919) & $\Gamma = 1.9 \pm 0.1$ & $(1.0 \pm 0.2) \times 10^{-5}$ & $(1.3 \pm 0.3) \times 10^{-5}$ \\
 & $E_c=4.0 \pm 1.2$\,GeV & &  \\
\hline
\celltspace P2 (PSR~J0537$-$6910 / N~157B) & $\Gamma = 2.2  \pm 0.1 $ & $(0.9 \pm 0.2) \times 10^{-5}$ & $(1.1 \pm 0.2) \times 10^{-5}$ \\
\hline
\celltspace P3 (Unassociated) & $\Gamma = 2.8 \pm 0.1$ & $(5.2 \pm 0.6) \times 10^{-6}$ & $(8.9 \pm 1.3) \times 10^{-6}$ \\
\hline
\celltspace P4 (N~132D) & $\Gamma = 1.4 \pm 0.3$ &$(1.9 \pm 1.4) \times 10^{-6}$  & $(1.9 \pm 1.4) \times 10^{-6}$ \\
\hline
\end{tabular}
\end{center}
From left to right, the columns list the source identifier used in this work and in parenthesis the established or possible counterpart (see text); spectral parameters, a power-law photon index for all sources plus the cutoff energy for P1; and the energy flux in 0.2--100\,GeV and extrapolated to the 0.1--100\,GeV range in MeV\,cm$^{-2}$\,s$^{-1}$ units.
\end{minipage}
\end{table*}

\begin{figure}[!t]
\begin{center}
\includegraphics[width= 8cm]{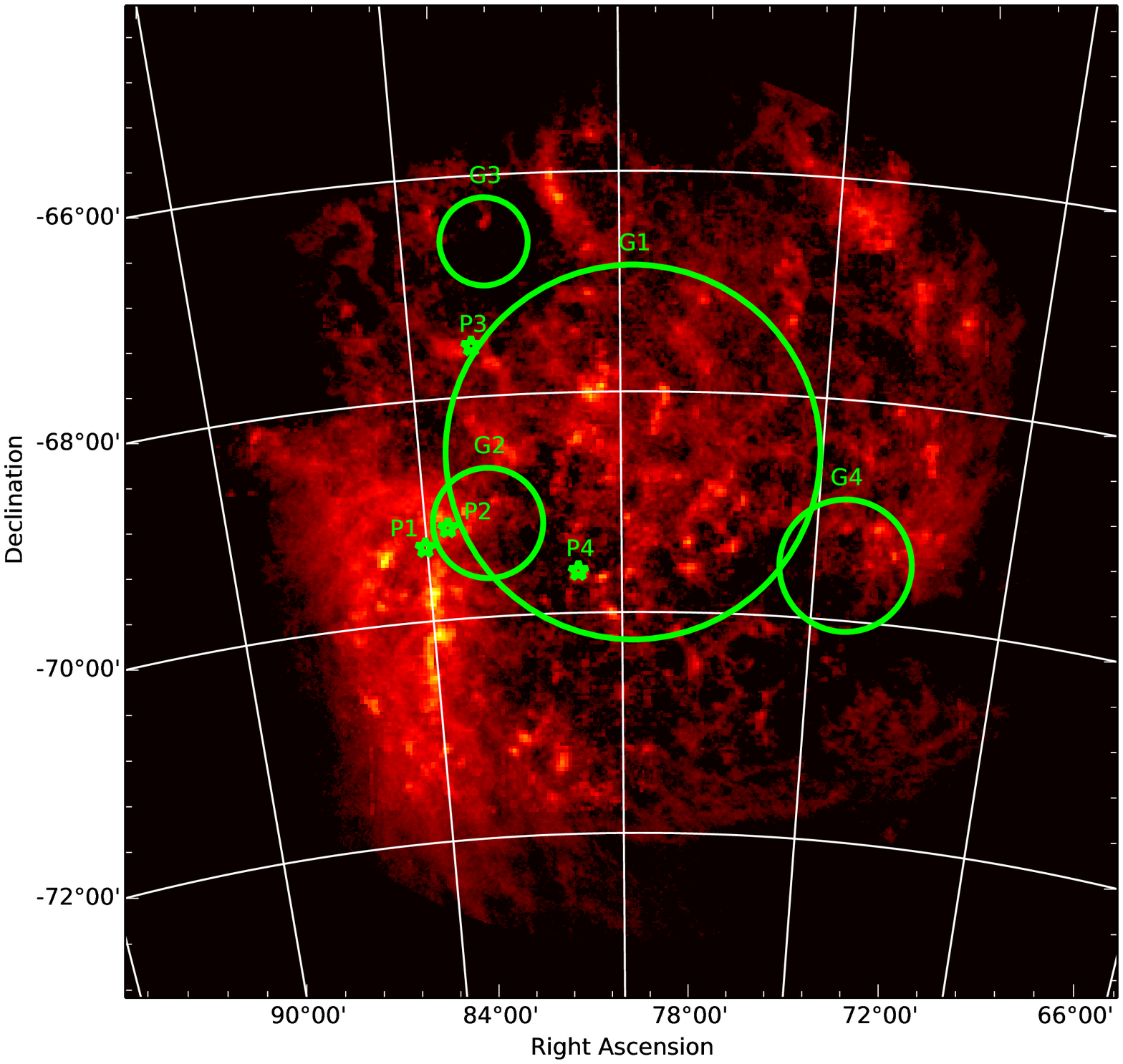}
\includegraphics[width= 8cm]{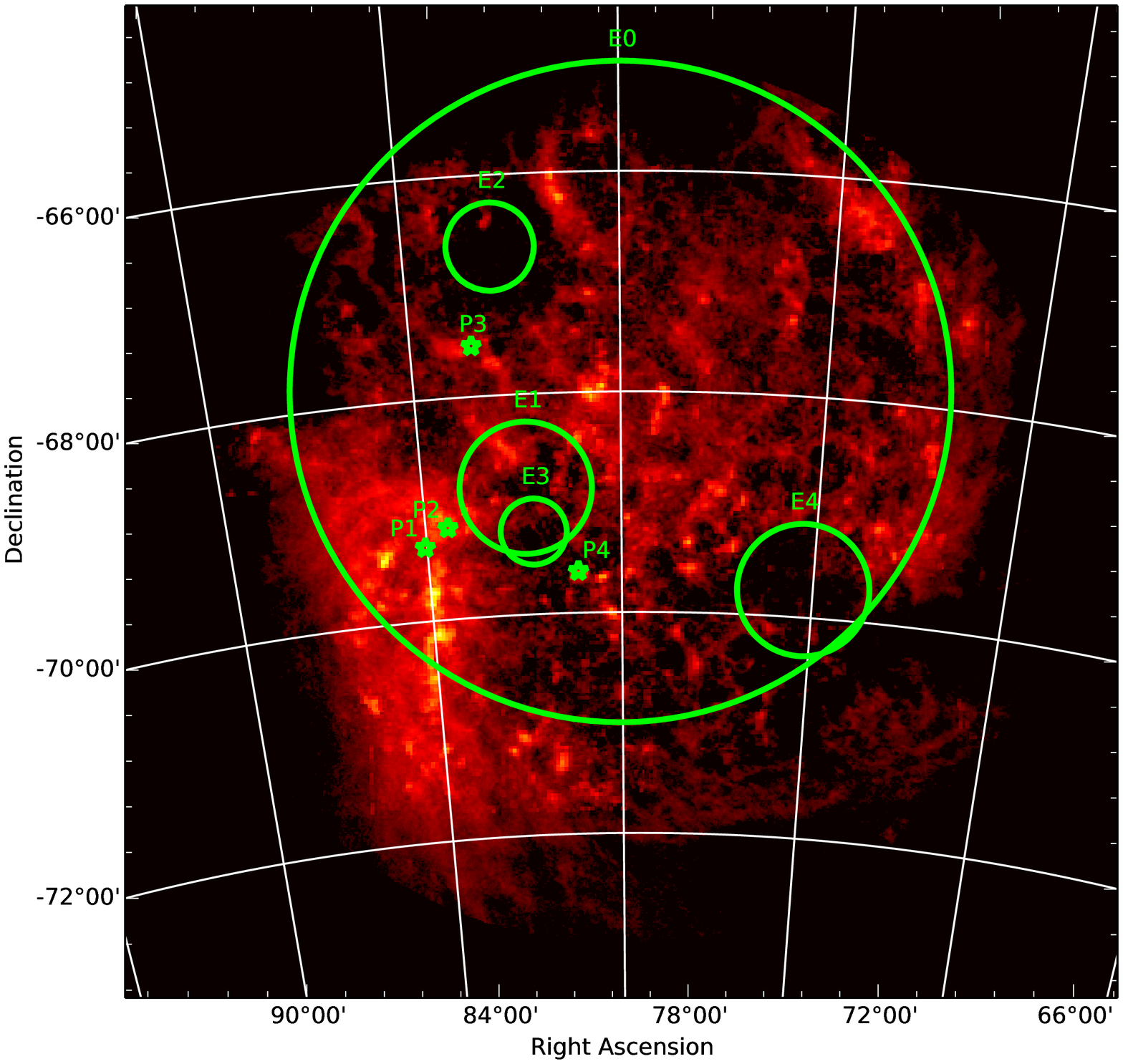}
\caption{Charts illustrating the layout of the model components. Point sources are indicated by green stars. The green circles correspond to the 1-$\sigma$ extent of the Gaussian emission components of the analytic model (top) and to the 1-$\sigma$ extent of the Gaussian emissivity components of the emissivity model (bottom). The background map is the total gas column density distribution in arbitrary units and a square-root scale.}
\label{fig_charts}
\end{center}
\end{figure}

As expected, this new approach results in a layout of the components which is very close to that of the analytic model. There are five extended emissivity components, labelled from E0 to E4, as listed in Table \ref{tab_emimodel} and illustrated in Fig. \ref{fig_charts}. However, two of them are adjacent and may well be one single component with a shape that is poorly fit by a 2D Gaussian function (components E1 and E3 in the bottom chart of Fig. \ref{fig_charts}). When we grouped the corresponding spatial templates into a single component associated with a power-law spectral model, we found that the fit was hardly degraded (variation in $\log \mathcal L < 1$). We therefore group the components E1 and E3 of the emissivity model together in the following, under the name E1+E3.

One advantage of this method is that the emissivity model retains the small-scale structure that the \gray\ emission may actually have, compared to the Gaussian morphologies of the analytic model.
The emissivity model provides a better fit than the analytic model, with a difference of 36 in $\log \mathcal L$ for an additional 2 degrees of freedom, and therefore is the primary model used in this work to represent the emission from the LMC.

\subsection{Template fitting}
\label{model_tpl}

In \citetalias{Abdo:2010d}, the emission model providing the highest likelihood was a map of the ionised gas column density across the LMC. With such templates from observations at other wavelengths, we can compare hypotheses about the spatial distribution of the \gray\ emission, which can provide clues about the origin of the emission. In \citetalias{Abdo:2010d}, this revealed a surprising correlation of the \gray\ emission with ionised gas, which traces the population of young and massive stars, and a poor correlation with total gas, as could have been expected if CRs were almost uniformly distributed in the LMC.

We revisited this result with the enlarged data set now available, and especially in light of the detection of point sources within the LMC. We performed fits of the ionised gas and total atomic+molecular+ionised gas column density distributions, using the same maps as in \citetalias{Abdo:2010d}.
For each map, we added as other components of the LMC emission model either \object{PSR~J0540$-$6919} alone (because it is the only firmly identified source), or all four point sources listed in Table \ref{tab_ptsrc}, with free spectral parameters. The resulting $\log \mathcal L$ are given in Table \ref{tab_allmodels}.

The ionised gas template still accounts better for the data than the total gas distribution. The impact of the spin temperature used in the preparation of the atomic gas column density map is negligible compared to the difference between the two templates. But the ionised gas template features a strong peak in the \object{30 Doradus} region that pushes sources P1 and P2 to negligible $TS$ values \citep[P1 in particular is pushed down to a flux level below the pulsed fraction of the signal, which is about 75\%; see][]{Fermi-collaboration:2015a}. 
The ionised gas template is therefore inappropriate to model the \object{PSR~J0540$-$6919} point source and the extended emission at the same time, while the total gas template provides much poorer fits to the data than the emissivity or analytic model (differences of $\sim$400 in $\log \mathcal L$, for $\sim$20 fewer degrees of freedom).

\subsection{Summary of all models}
\label{model_sum}

\begin{figure}[H]
\begin{center}
\includegraphics[width= 7.1cm]{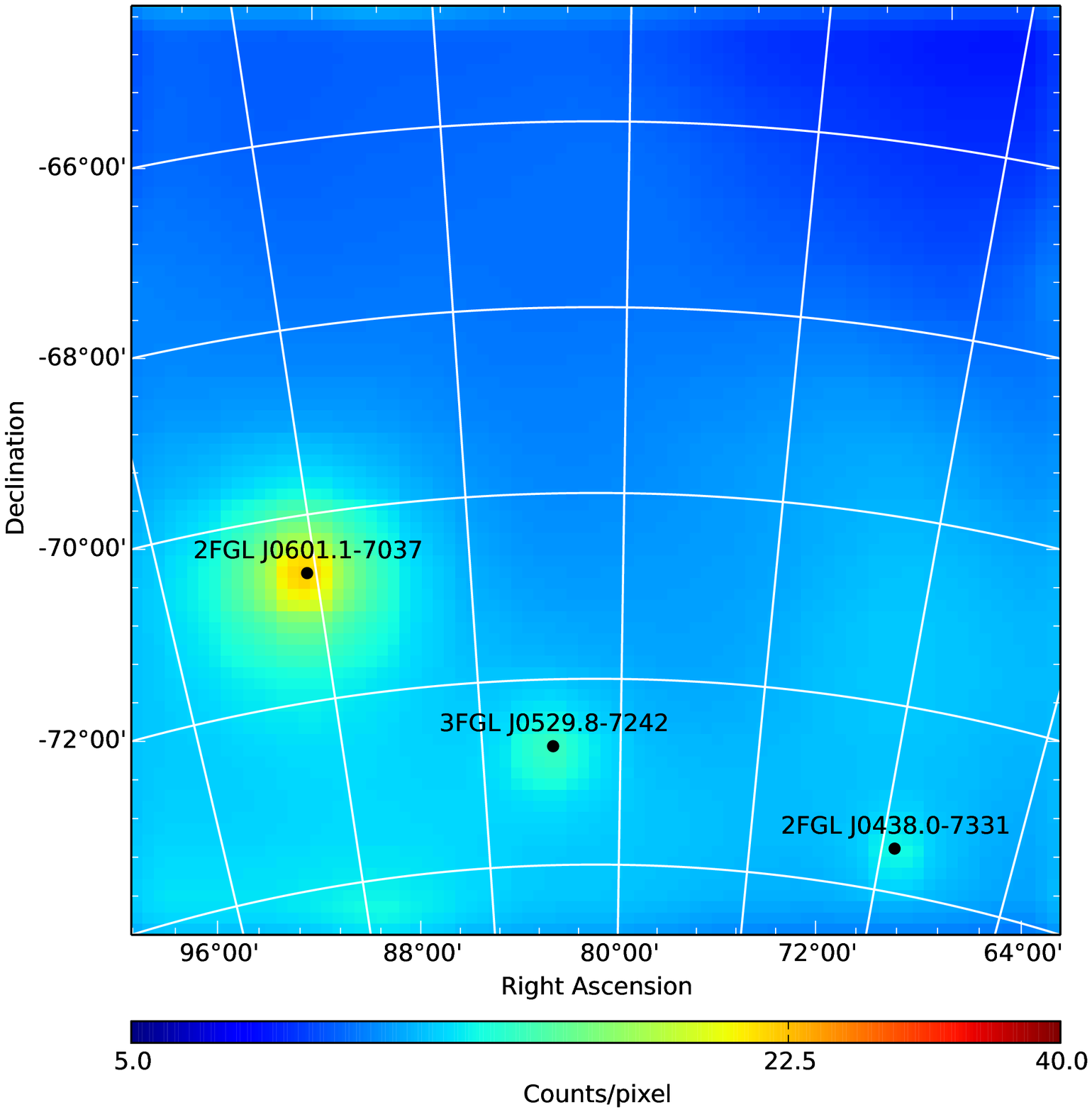}
\includegraphics[width= 7.1cm]{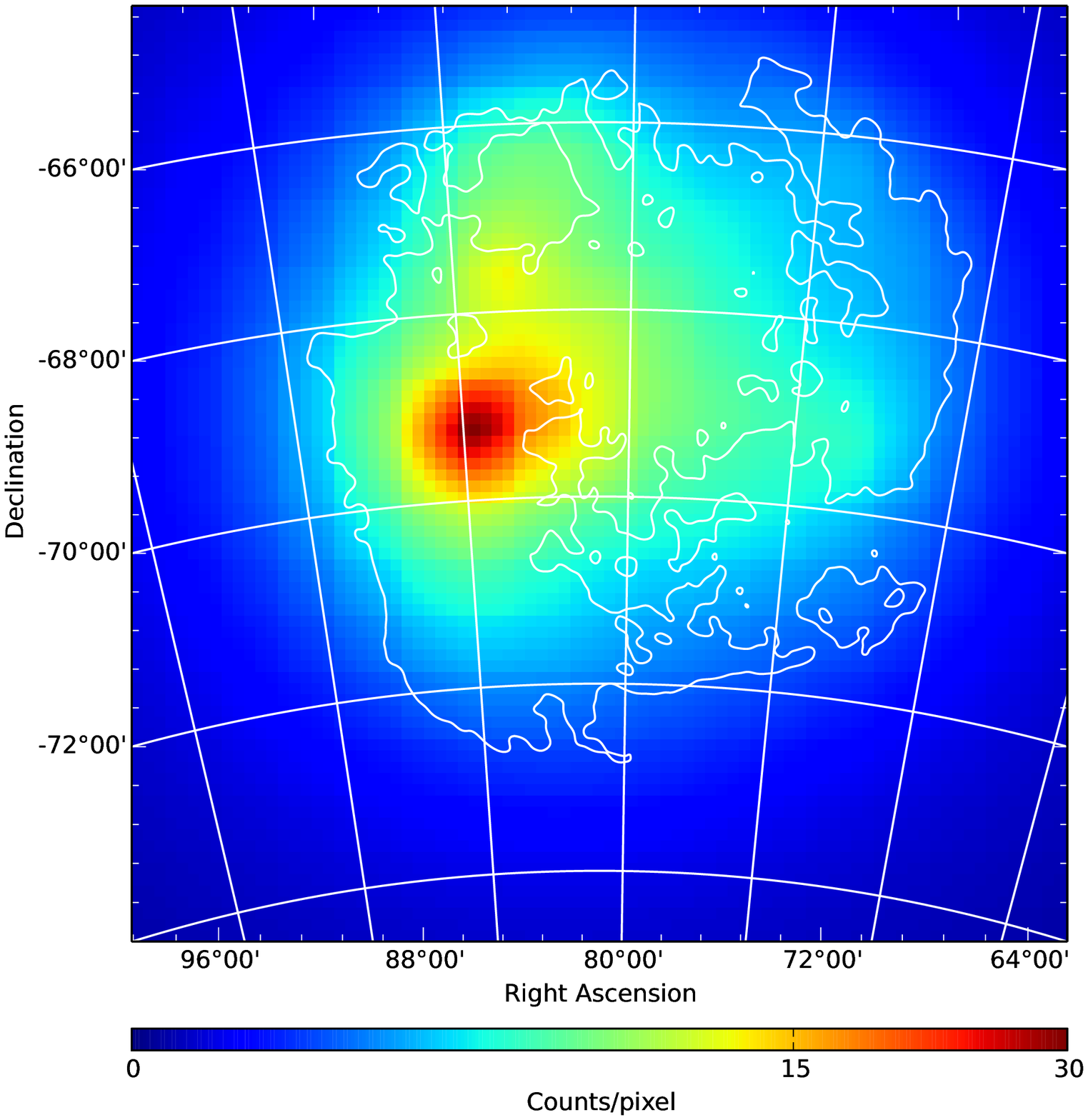}
\includegraphics[width= 7.1cm]{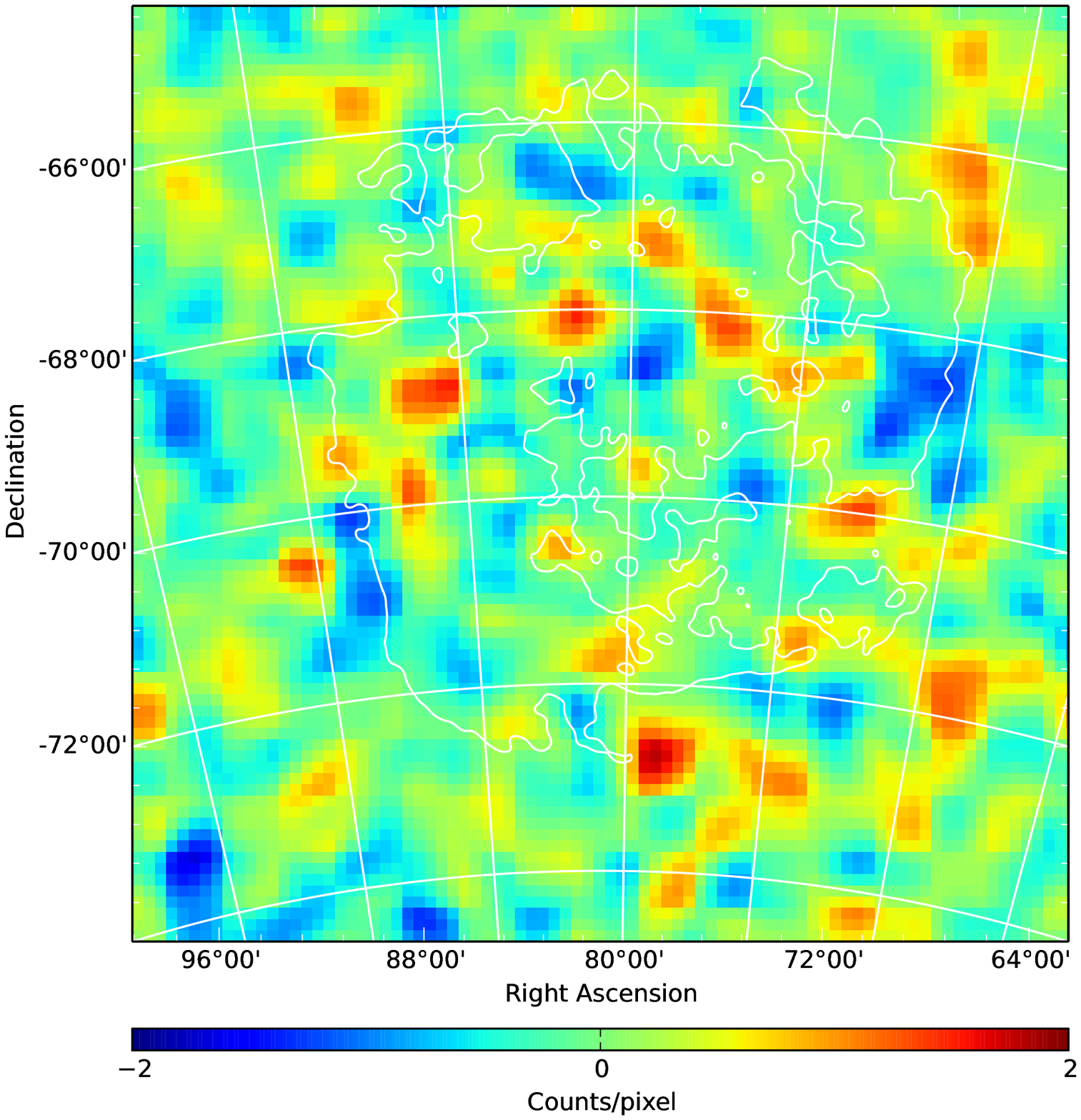}
\caption{Top and middle panels: model maps of the background components and LMC components of the emissivity model in the 0.2--100\,GeV band. Black dots indicate the positions of background sources in the field but outside the LMC boundaries (see Sect. \ref{model_bgd}). Both maps have $0.1\deg \times 0.1\deg$ pixels and were smoothed with a Gaussian kernel with $\sigma=0.2\deg$. Colours are displayed on a square-root scale. Bottom panel: residual counts map in the 0.2--100\,GeV range, after subtracting the fitted emissivity model. Colours are displayed on a linear scale. White lines are defined in Fig. \ref{fig_cntmap}.}
\label{fig_modmap}
\end{center}
\end{figure}

The various emission models tested are summarised in Table \ref{tab_allmodels}. The model yielding the highest likelihood is the emissivity model\footnote{The significance of the improvement over the analytic model or the best template and point sources combination cannot be quantified easily, however, since models are not nested \citep{Protassov:2002}.}.
The corresponding model maps in the 0.2--100\,GeV range for the background and LMC components are shown in the top and middle panels of Fig. \ref{fig_modmap}, respectively, and a residual counts map is plotted in the bottom panel of Fig. \ref{fig_modmap}. The distribution of residuals is satisfactory, with a mean close to zero and balanced negative and positive residuals. The strongest residual excess towards $(\alpha,\delta)=(79.0^\circ,-72.8^\circ)$ corresponds to a $TS$ of 17.

The outcome of the analysis is insensitive to the size of the ROI; we repeated the analysis for larger fields of up to $16\deg \times 16\deg$, and the significance of the various emission components as well as the normalisations of the isotropic and Galactic diffuse emission templates were not significantly altered. Since the latter is a global model and may be a source of systematics for the specific region of the LMC, we also ran an analysis where its different components (the various gas templates and the inverse-Compton model) were fitted independently; the fit was not significantly improved. Overall, the analysis is robust against the background model.

\section{Point-like sources}
\label{res_ptsrc}

This section is dedicated to the four point sources found within the LMC that are labelled P1, P2, P3, and P4 in Table \ref{tab_ptsrc}. As detailed below, P1 is unambiguously identified as pulsar \object{PSR~J0540$-$6919} through its characteristic pulsed \gray\ emission. For the other sources, possible associations with known objects are discussed, based essentially on their positions and spectra.

Figure \ref{fig_loca_ptsrc} shows the confidence regions for the localisation of point sources P2, P3, and P4 (P1 has its position fixed to that of the pulsar in optical), plotted on top of a SHASSA H$\alpha$ smoothed map\footnote{SHASSA stands for Southern H-Alpha Sky Survey Atlas; more information at http://amundsen.swarthmore.edu}. For each source, we overplotted the positions of known SNRs and pulsars in the field, according to the SIMBAD database. Figure \ref{fig_spec_ptsrc} shows the best-fit spectral shape for each source, together with spectral points or upper limits obtained from fits in five individual bins per decade in the 0.2--200\,GeV range\footnote{Spectral points were computed assuming a power-law with fixed photon index 2.1 within each bin; upper limits correspond to 95\% confidence level and were computed by searching for the flux yielding a variation of $\log \mathcal L$ by 1.35 from the optimum.}. Table \ref{tab_ptsrcspec} summarises the best-fit spectral parameters of each source and provides the energy flux in the 0.2--100\,GeV band and its extrapolation to the 0.1--100\,GeV band.

\newpage

\newpage

\subsection{P1: Pulsar PSR~J0540$-$6919}
\label{res_pt_psr0540}

Source P1 is detected with a $TS$ of 151 and found at a position that is consistent with that of pulsar \object{PSR~J0540$-$6919} (also known as PSR B0540-69). \object{PSR~J0540$-$6919} is a 50 ms pulsar associated with the $\sim$1100-year old SNR 0540$-$69.3 and PWN N~158A \citep{Seward:1984,Williams:2008,Brantseg:2014}. With a spin-down luminosity of $\dot E = 1.5 \times 10^{38}$\,erg\,s$^{-1}$, it has the third largest spin-down power of any pulsar currently known, right behind that of the Crab pulsar, $4.5 \times 10^{38}$\,erg\,s$^{-1}$ (see the ATNF catalogue\footnote{See \citet{Manchester:2005} and the ATNF Pulsar Catalog version 1.52 at http://www.atnf.csiro.au/people/pulsar/psrcat/.}).

The possible contribution of this exceptional pulsar to the observed \gray\ signal from the \object{30 Doradus} region was already considered in \citetalias{Abdo:2010d}, but could not be secured: characteric pulsations of the signal were not detected at a significant level, and there was some confusion in the morphology of the region regarding the presence of a point source in the more extended emission of the \object{30 Doradus} region. \object{PSR~J0540$-$6919} is now unambiguously detected in GeV \grays\ from its characteristic pulsations. It is the first \gray\ pulsar detected in a galaxy other than the Milky Way and the most distant known.
A dedicated paper \citep{Fermi-collaboration:2015a} reports more details about this new \gray\ pulsar, such as phase-aligned lightcurves in different bands and a complete spectral energy distribution.

For consistency with the other sources discussed below, we show in Fig. \ref{fig_spec_ptsrc} the spectrum of the source obtained from P7REP data. The best-fit spectral shape is a power law with exponential cutoff, with a statistical significance at the $4\sigma$ level for the cutoff.

\subsection{P2: Coincident with PSR~J0537$-$6910 / N~157B}
\label{res_pt_psr0537}

\begin{figure}[H]
\begin{center}
\includegraphics[width= 7.3cm]{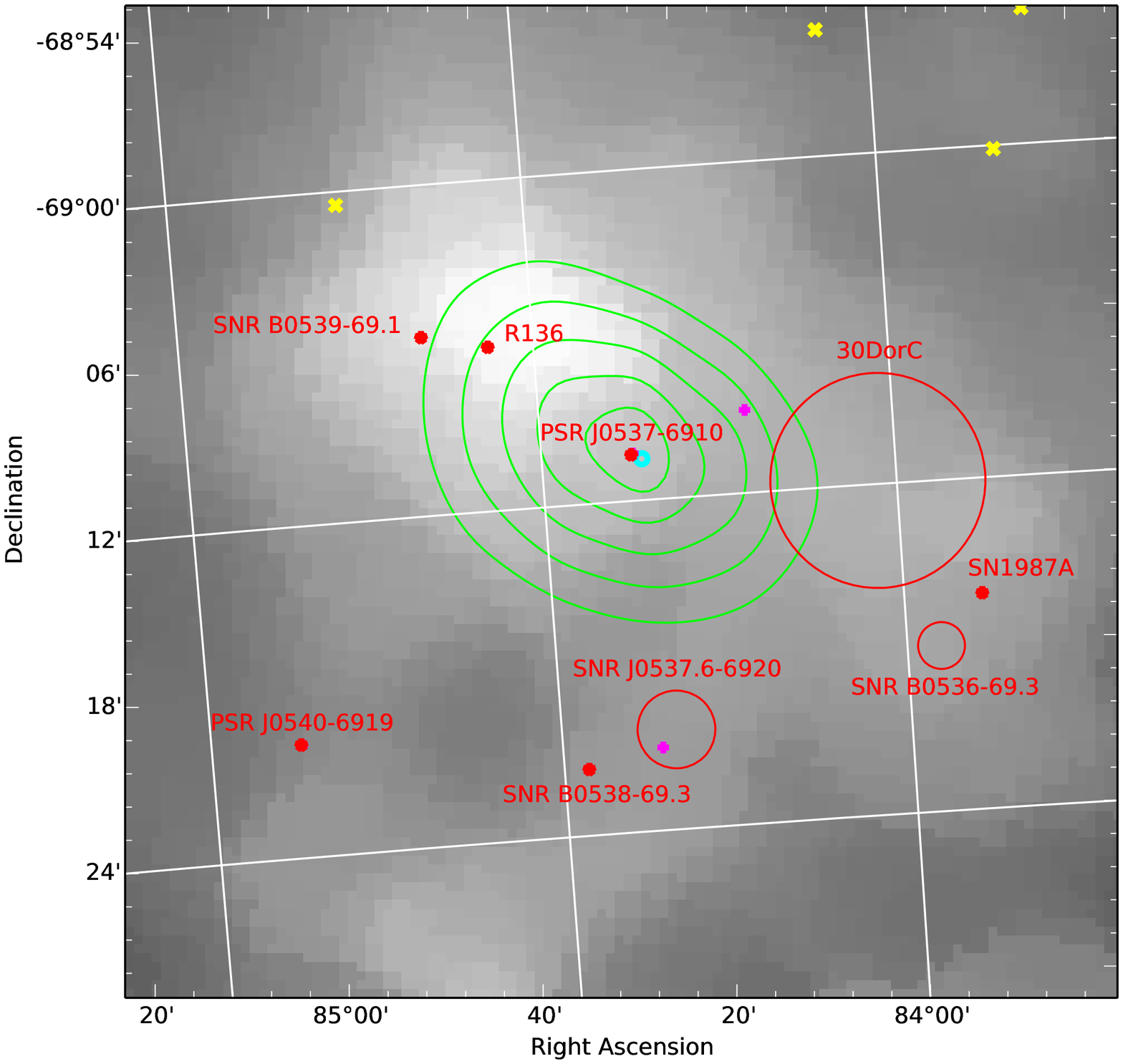}
\includegraphics[width= 7.3cm]{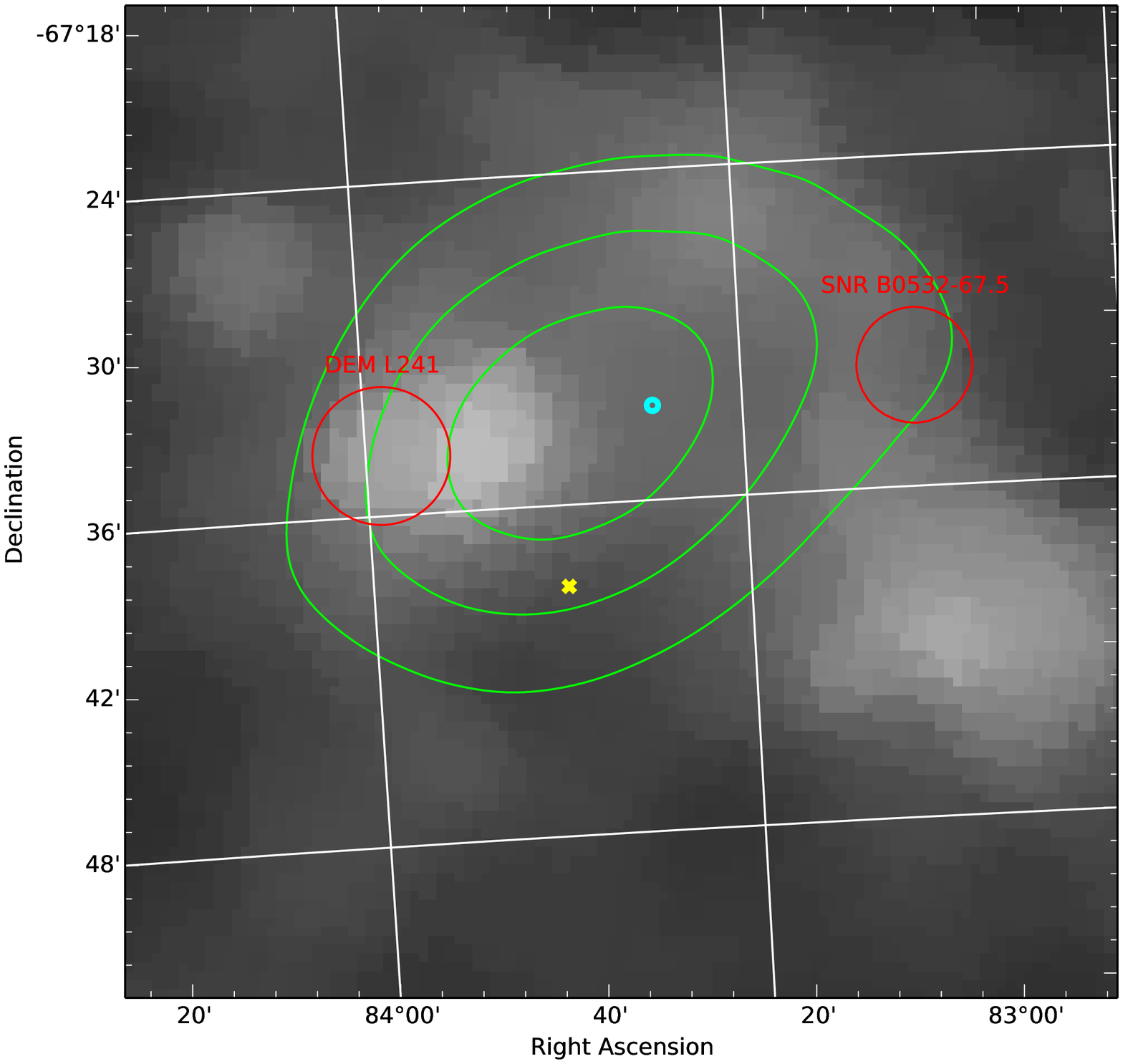}
\includegraphics[width= 7.3cm]{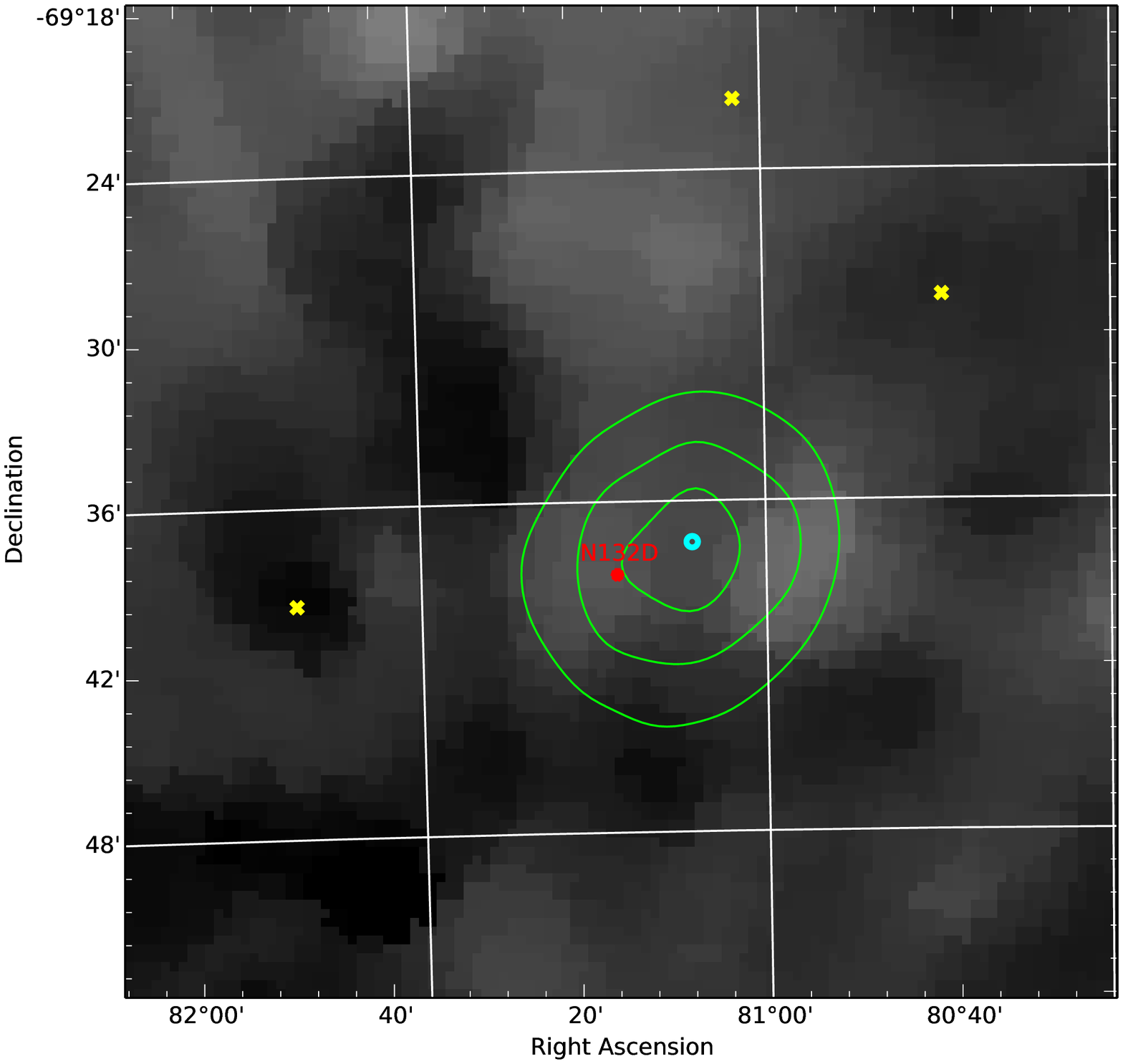}
\caption{Localization maps for point sources P2, P3, and P4 (from top to bottom; P1 is identified as PSR~J0540$-$6919 and thus not shown). Green contours correspond to 1, 2, and 3$\sigma$ confidence regions derived using the emissivity model (4 and 5$\sigma$ confidence regions were added to P2 to emphasise that it is not consistent with 30 Dor C); the cyan dots correspond to the best-fit positions derived using the analytic model. The background is a SHASSA H$\alpha$ smoothed map on a log scale. Overlaid as red dots and circles are the position of known SNRs, including the plerions associated with PSRs J0540$-$6919 and J0537$-$6910, together with other noteworthy objects such as SN1987A or \object{30 Doradus C}; magenta pluses indicate the positions of known pulsars, while yellow crosses indicate the positions of known active galactic nuclei.}
\label{fig_loca_ptsrc}
\end{center}
\end{figure}

Source P2 is detected with a $TS$ of 96 and found at a position which is consistent with that of pulsar \object{PSR~J0537$-$6910} (see Fig. \ref{fig_loca_ptsrc}). \object{PSR~J0537$-$6910} is a 16 ms pulsar associated with the $\sim$5000-year old SNR 0537.8$-$6910 and PWN \object{N~157B}\footnote{The complete nomenclature is LHA 120-N~157B but we use N~157B for short. The same remark applies to N~158A, \object{N~132D}, N~11, and other sources discussed further down.}. The whole system presumably results from the explosion of a $\sim$25\msol\ O8-O9 star and is likely associated with the OB association LH99 \citep{Chen:2006,Micelotta:2009}. The pulsar was discovered from X-ray observations \citep{Marshall:1998} and is the most rapidly spinning and the most powerful young pulsar known, with a spin-down luminosity of $\dot E = 4.9 \times 10^{38}$\,erg\,s$^{-1}$.

PWN \object{N~157B} was detected in TeV \grays\ with H.E.S.S. \citep{Abramowski:2012,HESS-collaboration:2015}. It is the first and only PWN detected outside of the Milky Way at these photon energies, which can be explained by the enormous spin-down power of the pulsar combined with a rich infrared photon field for inverse-Compton scattering. The emission in the {\em Fermi}-LAT range is observed at a level of $\sim$ 2--3 $\times 10^{-12}$\,erg\,cm$^{-2}$\,s$^{-1}$, quite similar to that of the TeV emission detected with H.E.S.S.. 

\begin{figure}[H]
\begin{center}
\includegraphics[width= 7.6cm]{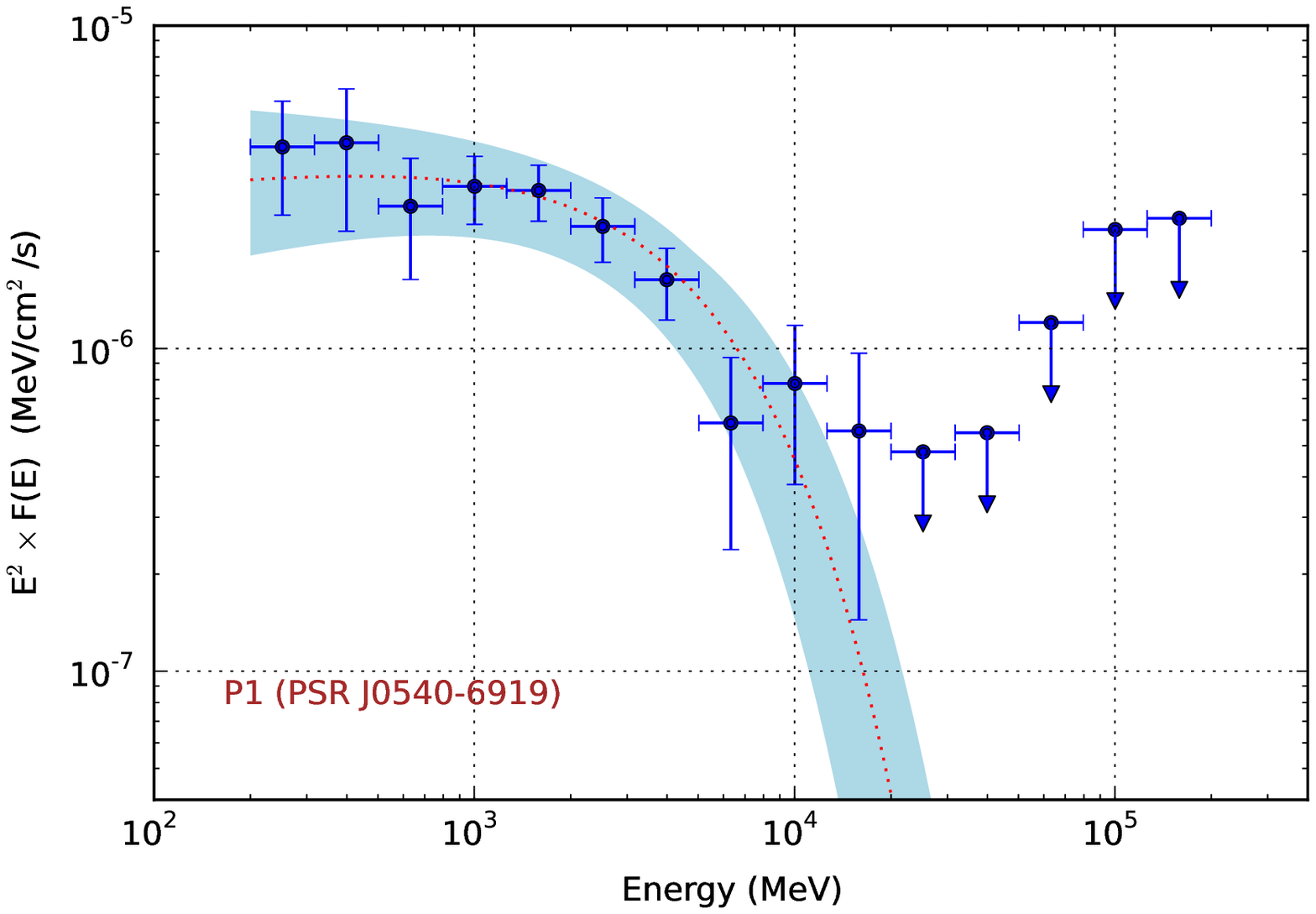}
\includegraphics[width= 7.6cm]{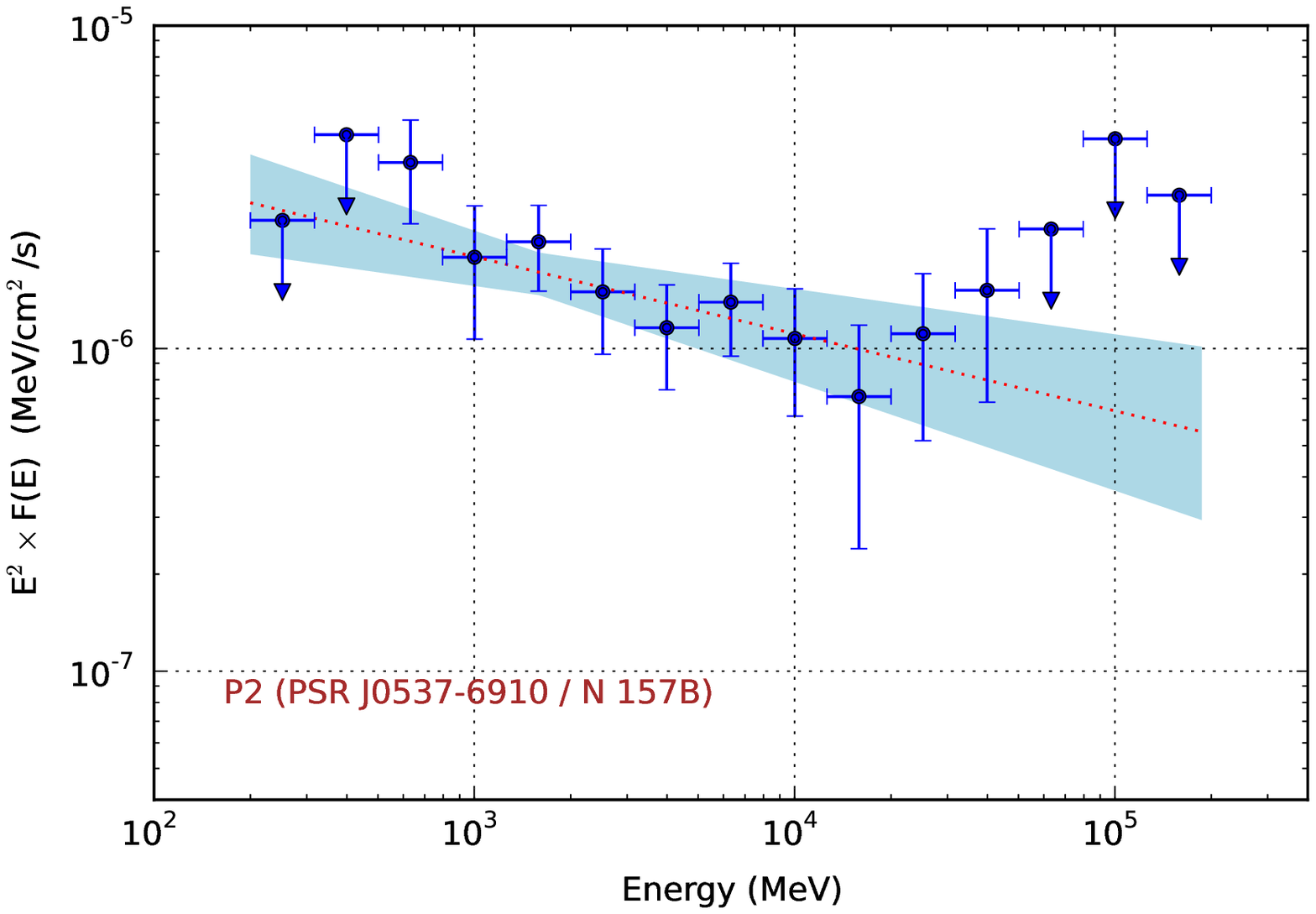}
\includegraphics[width= 7.6cm]{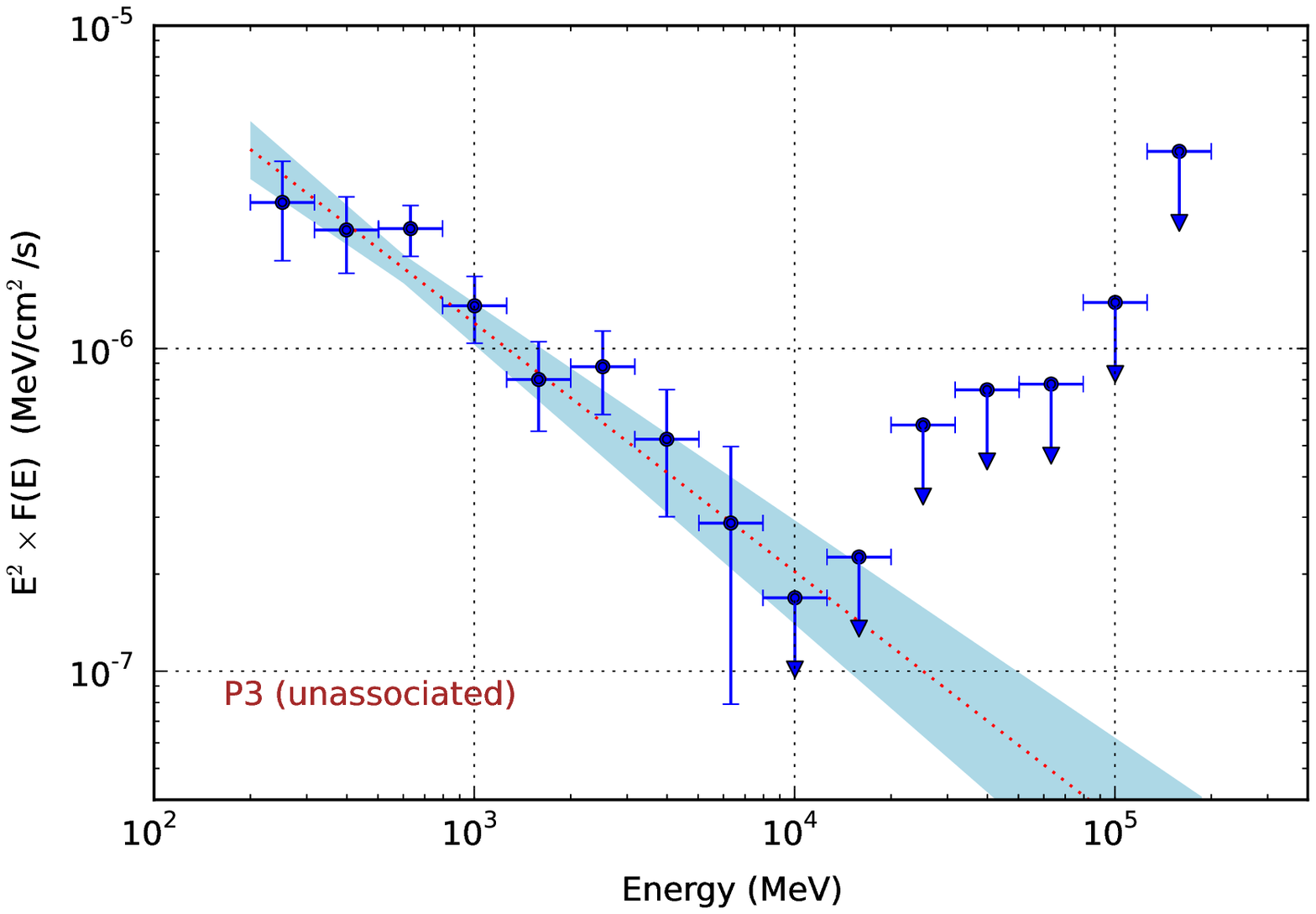}
\includegraphics[width= 7.6cm]{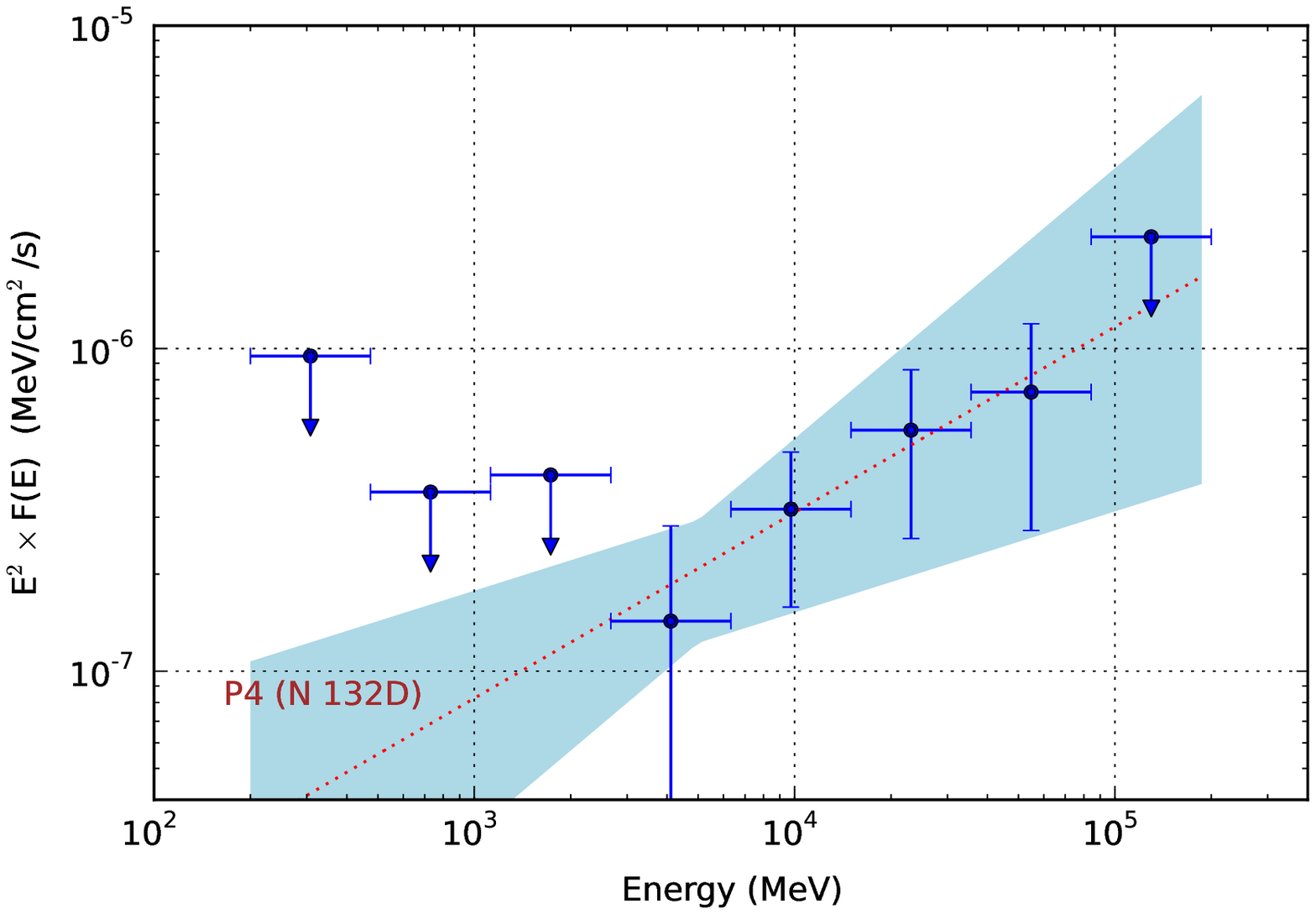}
\caption{Spectra of the four point-like sources found in the LMC, from top to bottom P1 to P4, as listed in Table \ref{tab_ptsrc}. We show spectral points obtained by fits in individual bins, together with the best-fit spectral model from a binned likelihood analysis over the full energy range. The spectrum of the weakest source P4 features coarser binning in energy. Upper limits (arrows) correspond to a 95\% confidence level.}
\label{fig_spec_ptsrc}
\end{center}
\end{figure}

As shown in Fig. \ref{fig_spec_ptsrc}, however, the spectrum of the source is rather flat over the 0.5--100\,GeV range, with a photon index of 2.2$\pm$0.1. The origin of the emission is unclear. It may come from the pulsar, the SNR, the PWN, or any combination of those in unknown proportions. The nearly flat or even declining spectrum seems hard to reconcile with emission from a PWN alone: in a typical scenario, emission from a PWN in the {\em Fermi}-LAT range is expected to have a hard spectrum as it probes the low-energy side of an inverse-Compton hump peaking at $\sim$TeV energies. In that respect, the spectral model considered in \citet{HESS-collaboration:2015} for the TeV data does not match our GeV observations at all. This situation is somewhat reminiscent of the Vela X PWN, for which GeV and TeV emissions at similar levels were postulated to arise from two different populations of emitting particles \citep{Abdo:2010j}. The observed emission may also result from the superposition of a pulsar's curved spectrum peaking at $\sim$1\,GeV and a PWN's spectrum ramping up and dominating at $\sim$10\,GeV. The detection of pulsations from the source and an analysis of its off-pulse emission would have been key to establish such a scenario. We searched for pulsations in the \gray\ signal in the same way as for \object{PSR~J0540$-$6919} \citep{Fermi-collaboration:2015a}, using an ephemeris for \object{PSR~J0537$-$6910} based on RXTE observations covering the first 3.5 years of the {\em Fermi} mission. No evidence for \gray\ pulsations was found, with a pulsation significance below 1$\sigma$. On longer time scales, the source shows no evidence of variability down to a monthly basis.

With a total luminosity of $5.2 \times 10^{36}$\,erg\,s$^{-1}$ in the 0.1--100\,GeV range, this source is remarkable. Further studies involving the complete picture of the plerionic system \object{N~157B} are required to assess the possible origin of the GeV emission. It is clear, however, that the accuracy of the localisation allows us to exclude superbubble \object{30 Doradus C} as the main contributor to the emission (see Fig. \ref{fig_loca_ptsrc}). The latter was detected in TeV \grays\ with H.E.S.S. \citep{HESS-collaboration:2015} and might have had a GeV counterpart detectable with {\em Fermi}-LAT. An upper limit on the emission from this object is given below, in Sect. \ref{res_uls}.

\subsection{P3: Unassociated source}
\label{res_pt_deml241}

Source P3 is detected with a $TS$ of 116 and found at a position which is consistent with that of HII regions \object{NGC 2029}/\object{NGC 2032}. Supernova remnant \object{DEM L 241} and Seyfert 1 galaxy \object{2E 1445} are also in the field but they lie at a greater distance from the best-fit position, at the edge of the $2\sigma$ confidence region.

The source shows no evidence of variability down to a monthly basis. The most notable feature of this source is its very soft power-law spectrum, with a photon index of $2.8 \pm0.1$. If source P3 is a background active galactic nucleus, such a soft spectrum tends to favour a flat spectrum radio quasar type over a BL Lacertae type \citep{Ackermann:2011c}.

\subsection{P4: Coincident with SNR N~132D}
\label{res_pt_n132d}

Source P4 is detected with a $TS$ of 28 and found at a position which is marginally consistent with that of SNR \object{N~132D}, just outside the $1\sigma$ confidence region (see Fig. \ref{fig_loca_ptsrc}). Supernova remnant \object{N~132D} belongs to the rare class of O-rich remnants, together with objects such as Cassiopeia A, Puppis A, or SNR 0540$-$69.3, and it has a kinematic age of 2500\,yr \citep[][and references therein]{Vogt:2011}. Its chemical composition indicates that \object{N~132D} is the remnant of the core-collapse supernova of a progenitor star with initial mass in the $\sim$35--75\,M$_\odot$ range \citep{France:2009}. The shock wave in the remnant is interacting with surrounding material, and \object{N~132D} is the brightest remnant of the LMC in X-rays \citep{Borkowski:2007}. 

The association of our weak GeV source with \object{N~132D} is supported by the recent marginal detection using H.E.S.S. of a TeV source spatially coincident with \object{N~132D} \citep[with a sub-arcmin localisation accuracy compared to the few arcmin allowed by {\em Fermi}-LAT for this source; see][]{HESS-collaboration:2015}. In the {\em Fermi}-LAT range, the source has a hard spectrum with photon index $1.4 \pm0.3$ and a differential flux of $\sim 10^{-6}$\,MeV\,cm$^{-2}$\,s$^{-1}$ at 100\,GeV. In the H.E.S.S. band, the differential flux is $\sim 2 \times 10^{-7}$\,MeV\,cm$^{-2}$\,s$^{-1}$ at 1\,TeV, with a photon index 2.4$\pm$0.3.  If the \gray\ source is \object{N~132D}, its spectral energy distribution in the \gray\ domain may have a peak around 100\,GeV, possibly a broad one given the uncertainty on the spectral index in the 1--100\,GeV range. The spectrum presented here is a factor $\sim5$ above the hadronic model considered in \citet{HESS-collaboration:2015}, which may be accommodated by a modest tuning of the model parameters; in contrast, the leptonic model seems excluded because of a much lower predicted flux and possibly also an inconsistent spectral slope.

\subsection{Systematic uncertainties}
\label{res_unc}

The properties of the four point sources discussed above may be affected by systematic errors, introduced by selecting a particular method to build the complete LMC emission model. As exposed in Sect. \ref{model}, two models provide a good fit to the data: the emissivity model, which yields the highest likelihood and is the reference one, and the analytic model, which gives a slightly lower likelihood but can be considered as a competitive alternative. The other models tested in this work were far from equaling these two (see Table \ref{tab_allmodels}). 

Interestingly, the 2D Gaussian extended components of the analytic model have no fine structures compared to the extended components of the emissivity model, which were built from the gas column density distribution. The analytic model can therefore be used to test whether such fine structures can bias the inferred properties of point sources.

When using the analytic model, the spectral parameters of the four point sources within the LMC turned out to be consistent within 1$\sigma$ statistical uncertainties with those obtained with the emissivity model. The differences in the mean values of photon and energy fluxes of point sources between the
two models were found to be $<50$\% of the statistical uncertainties.

The inferred properties of the sources may also be affected by uncertainties in the characterisation of the LAT instrument. According to \citet{Ackermann:2012c}, the magnitude of the systematic error arising from uncertainties in the effective area, point spread function, and energy dispersion, is about 0.1 on the spectral index and about 10\% on the photon and energy fluxes (for the analysis of a not-too-hard point source in the 0.1--100\,GeV range). These systematic uncertainties are therefore of the same order as the statistical uncertainties on spectral indices and fluxes for the three brightest point sources in the LMC.

\subsection{Upper limits on selected objects}
\label{res_uls}

Some particularly interesting LMC objects that were or could have been expected to be \gray\ sources were undetected in the {\em Fermi}-LAT observations we analysed. This is the case of superbubble \object{30 Doradus C} \citep[a rare superbubble with non-thermal emission, recently detected at TeV energies with H.E.S.S.; see][]{HESS-collaboration:2015,Yamaguchi:2010}, or SN 1987A \citep[the closest supernova in modern times, predicted to be accelerating particles and to be increasingly shining in \grays; see][]{Berezhko:2011}. Another possible source of GeV emission was N~11, the second most active star-forming region of the LMC after \object{30 Doradus}, and maybe an older version of it \citep{Rosado:1996,Walborn:1992}. The LMC also harbours many SNRs that are very well studied in X-rays and at radio wavelengths, and for which constraints on the \gray\ emission might be useful.

\newpage
\newpage

\begin{table}[!t]
\begin{center}
\caption{Upper limits on undetected sources}
\label{tab_uls}
\begin{tabular}{|c|c|c|}
\hline
\celltspace Source & $s=2.0$ & $s=2.5$ \cellbspace \\
\hline
\celltspace SN~1987A & $7.8 \times 10^{-7}$ & $6.6 \times 10^{-7}$ \\
\hline
\celltspace B0505-67.9 (DEM L71) & $5.9 \times 10^{-7}$ & $6.1 \times 10^{-7}$  \\
\hline
\celltspace B0506-68.0 (N~23) & $6.7 \times 10^{-7}$ & $7.0 \times 10^{-7}$  \\
\hline
\celltspace B0509-68.7 & $3.2 \times 10^{-7}$ & $3.8 \times 10^{-7}$  \\
\hline
\celltspace B0509-67.5 & $1.3 \times 10^{-7}$ & $1.4 \times 10^{-7}$  \\
\hline
\celltspace B0519-69.0 & $5.6 \times 10^{-7}$ & $5.9 \times 10^{-7}$  \\
\hline
\celltspace B0525-66.0 (N~49B) & $3.1 \times 10^{-7}$ & $3.9 \times 10^{-7}$  \\
\hline
\celltspace B0525-66.1 (N~49) & $3.1 \times 10^{-7}$ & $3.9 \times 10^{-7}$  \\
\hline
\celltspace B0534-69.9 & $3.2 \times 10^{-7}$ &  $2.7 \times 10^{-7}$ \\
\hline
\celltspace B0535-66.0 (N~63A) & $4.8 \times 10^{-7}$ &  $4.8 \times 10^{-7}$ \\
\hline
\celltspace B0548-70.4 & $1.6 \times 10^{-7}$ & $2.0 \times 10^{-7}$  \\
\hline
\celltspace 30 Doradus C & $1.1 \times 10^{-6}$ &  $8.9 \times 10^{-7}$ \\
\hline
\celltspace N~11 & $7.1 \times 10^{-7}$ & $7.1 \times 10^{-7}$  \\
\hline
\end{tabular}
\end{center}
Upper limits correspond to a 95\% confidence level and were computed under the assumption of a power-law spectrum. Column 1 lists the sources for which upper limits were computed, and Cols. 2 and 3 give the flux limits in units of MeV\,cm$^{-2}$\,s$^{-1}$ over the 1--10\,GeV interval for two power-law photon indices. All SNRs were modelled as point sources, while N~11 and 30 Dor C were modelled as 2D Gaussian intensity distributions with $\sigma=0.4\deg$ and $\sigma=0.1\deg$ , respectively.
\end{table}

Based on the emissivity model described previously, we derived upper limits on the 1--10\,GeV flux from the three objects mentioned above and the ten brightest remnants in X-rays, according to the MCSNR database\footnote{www.mcsnr.org}. These are summarised in Table \ref{tab_uls} under two assumptions for the photon index over 1--10\,GeV. 

For \object{SN1987A}, our upper limit on the 1--10\,GeV flux is not constraining when compared to the predictions made by \citet{Berezhko:2011}; our limit is a factor of 3 above their predicted flux for 2010 and remains above the maximum flux predicted for 2030. For 30 Dor C, our upper limit is not constraining either when compared to the \gray\ emission models considered in \citet{HESS-collaboration:2015}; our limit is a factor of 6 above the most optimistic prediction of the hadronic model.

\begin{table}[!t]
\begin{center}
\caption{Comparison of spectral models for component E0}
\label{tab_compspec}
\begin{tabular}{|c|c|c|}
\hline
\celltspace Model &  $\log \mathcal L$ & N$_{\textrm{dof}}$ \cellbspace \\
\hline
\celltspace Power law & -102523.4 & 2 \\
\hline
\celltspace Power law with cutoff & -102499.6 & 3 \\
\hline
\celltspace Broken power law & -102497.3 & 4 \\
\hline
\celltspace Log parabola & -102496.3 & 3 \\
\hline
\celltspace Tabulated & -102495.7 & 1 \\
\hline
\end{tabular}
\end{center}
$\log \mathcal L$ values correspond to a fit over 0.2--10\,GeV with 17 logarithmic bins (because component E0 is not significantly detected above 10\,GeV), hence the difference with values in Table \ref{tab_allmodels}.
\end{table}

\newpage
\newpage

\section{Extended sources}
\label{res_extsrc}

This section is dedicated to the extended sources found in the modelling of the \gray\ emission from the LMC. We used as reference the emissivity model and the components labelled E0, E1, E2, E3, and E4 in Table \ref{tab_emimodel}. We recall that components E1 and E3 can be grouped in the fit of the model, because leaving them independent does not provide a significant improvement; however, they need to be considered separately in some parts of the discussion below.

We first discuss the localisation and spectrum of each extended source. Figure \ref{fig_loca_ext} illustrates the layout of each region associated with extended \gray\ emission. Figure \ref{fig_spec_ext} shows the best-fit spectral shape for each component, together with spectral points and upper limits derived over the 0.2--200\,GeV range \footnote{Data points and upper limits computed as described in footnote 5.}. Table \ref{tab_extspec} summarises the best-fit spectral parameters of each extended emission component and provides the energy flux in the 0.2--100\,GeV band and its extrapolation to the 0.1--100\,GeV band.

\subsection{Localisation}
\label{res_ext_loca}

The large-scale component labelled E0 is by construction emission coming from the whole LMC disk. A large part of it probably results from a large-scale population of CRs accumulated over long time scales and interacting with interstellar gas (see Sect. \ref{res_ext_emiss}). The origin of the emission from the three to four smaller-size extended components is more uncertain. It is not correlated with prominent features of the gas column density distribution, which suggests at least two possibilities: these relatively gas-poor regions harbour an overdensity of CRs as a result of a local concentration of CR sources, or they host a population of many discrete \gray\ emitters with individual fluxes below the detection threshold. To explore both options, we searched for correlation with structures or objects in the LMC. In Fig. \ref{fig_loca_ext}, the location and size of emissivity components (from the emissivity model) and emission components (from the analytic model) are plotted on top of a SHASSA H$\alpha$ smoothed map, which indirectly reveals the distribution of young stars. We also overlaid the distributions of various objects: pulsars, which are the dominant class of GeV sources in the Galaxy; Wolf-Rayet stars, which reveal where the most massive and shortest-lived stars are located; high-mass X-ray binaries (HMXBs); stars of spectral type from B0 to B3, which offer a longer perspective on recent star formation because their typical lifetime is a few 10\,Myr; supergiant shells (approximated as circles), which are structures with typical ages 5-10\,Myr \citep{Kim:1999}, testifying to strong releases of kinetic energy in the ISM, from which CRs could be accelerated and confined. 

\begin{table*}[!t]
\begin{minipage}[][5.4cm][c]{\textwidth}
\begin{center}
\caption{Best-fit spectral parameters of the extended components}
\label{tab_extspec}
\begin{tabular}{|c|c|c|c|}
\hline
\celltspace Source & Parameters & F$_{200}$ & F$_{100}$ \cellbspace \\
\hline
\celltspace E0 (large-scale disk) & $a=2.17 \pm 0.04$  & $(5.9 \pm 0.3) \times 10^{-5}$ &  $(6.7 \pm 0.4) \times 10^{-5}$ \\
& $b=0.22 \pm 0.02$ &  & \\
& $E_b=1.0$\,GeV &  & \\
\hline
\celltspace E1+E3 (west of 30 Doradus) & $\Gamma = 2.13 \pm 0.05$ & $(2.1 \pm 0.2) \times 10^{-5}$ & $(2.5 \pm 0.2) \times 10^{-5}$ \\
\hline
\celltspace E2 (northern LMC) & $\Gamma = 2.00 \pm 0.06$ & $(1.3 \pm 0.2) \times 10^{-5}$ & $(1.4 \pm 0.2) \times 10^{-5}$ \\
\hline
\celltspace E4 (western LMC) & $\Gamma = 2.13 \pm 0.07$ & $(0.9 \pm 0.1) \times 10^{-5}$ & $(1.1 \pm 0.1) \times 10^{-5}$ \\
\hline
\end{tabular}
\end{center}
Note to the Table: From left to right, columns are: the source identifier used in this work; the spectral parameters, a power-law photon index for the 
last three sources, and log-parabola parameters for E0; the energy flux in 0.2--100\,GeV and extrapolated to the 0.1--100\,GeV range in MeV\,cm$^{-2}$\,s$^{-1}$ units.
\end{minipage}
\end{table*}

No obvious correlation appears to be valid for all three regions at the same time. The most striking feature is the coincidence of emissivity component E2 with a large void delimited by a supergiant shell. The cavity is relatively empty of objects, at least those considered above; young stars of the Wolf-Rayet class to early B spectral type tend to be found on the periphery \citep[the cluster of B stars on the southern edge of the cavity is NGC 2004; see][]{Niederhofer:2015}. This also seems to be the case for emissivity component E4, although less remarkably so. The situation is less clear
for emissivity components E1 and E3 owing to the crowding of the region. Most young stars are still found outside the 1-$\sigma$ contour of the emissivity component; in particular, the rich \object{30 Doradus} area and the surrounding HII regions are clearly offset
from the core of the emissivity component. Maybe the most interesting feature is the correlation with the highest concentration of supergiant shells in the LMC, which testifies to intense star formation activity over the past 5--10\,Myr. The absence of conspicuous voids associated with some of these supergiant shells may also indicate that projection effects complicate such a search for spatial correlation.

\subsection{Spectra}
\label{res_ext_spec}

\begin{figure}[H]
\begin{center}
\vspace{1.5mm}
\includegraphics[width= 7.8cm]{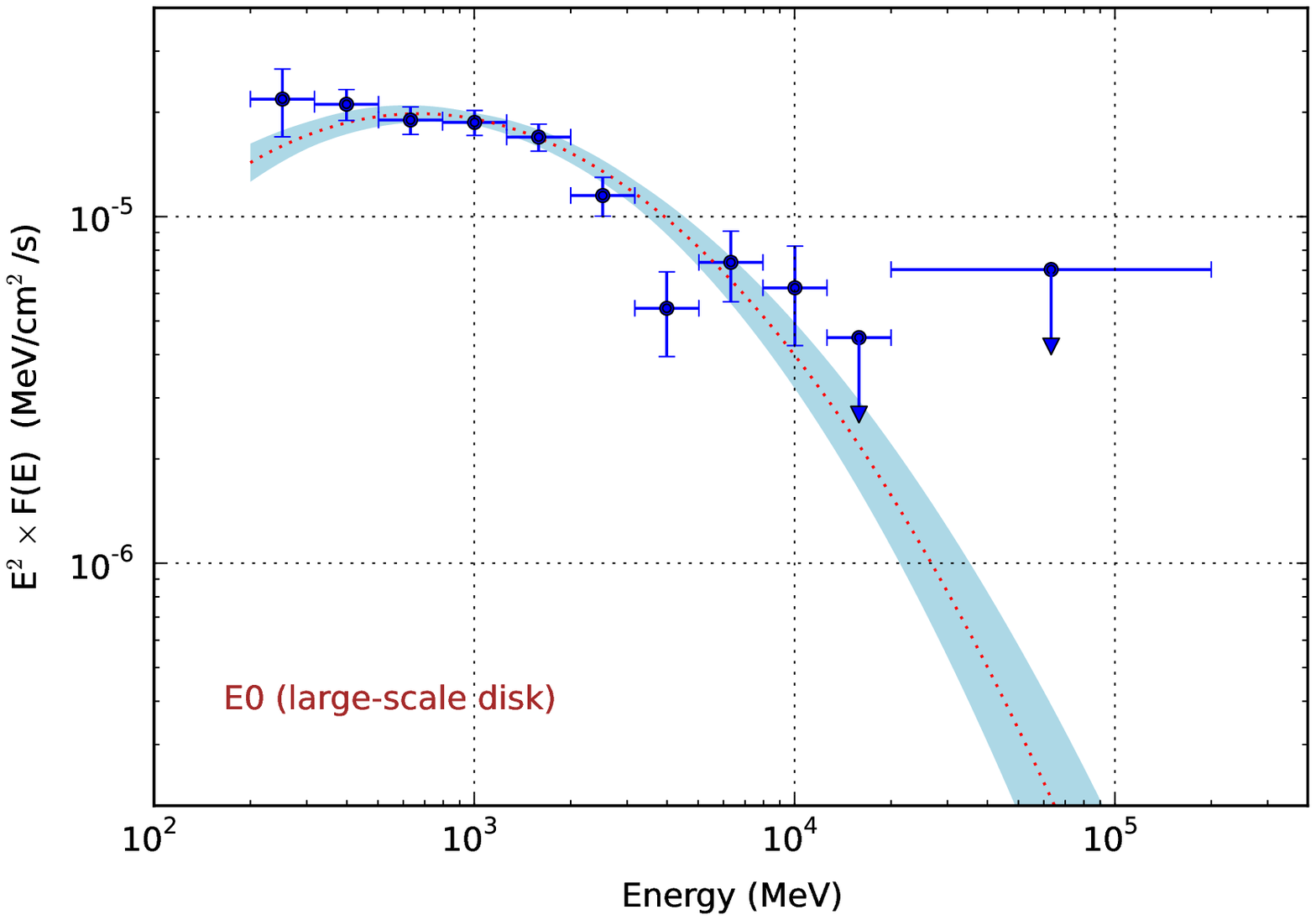}
\includegraphics[width= 7.8cm]{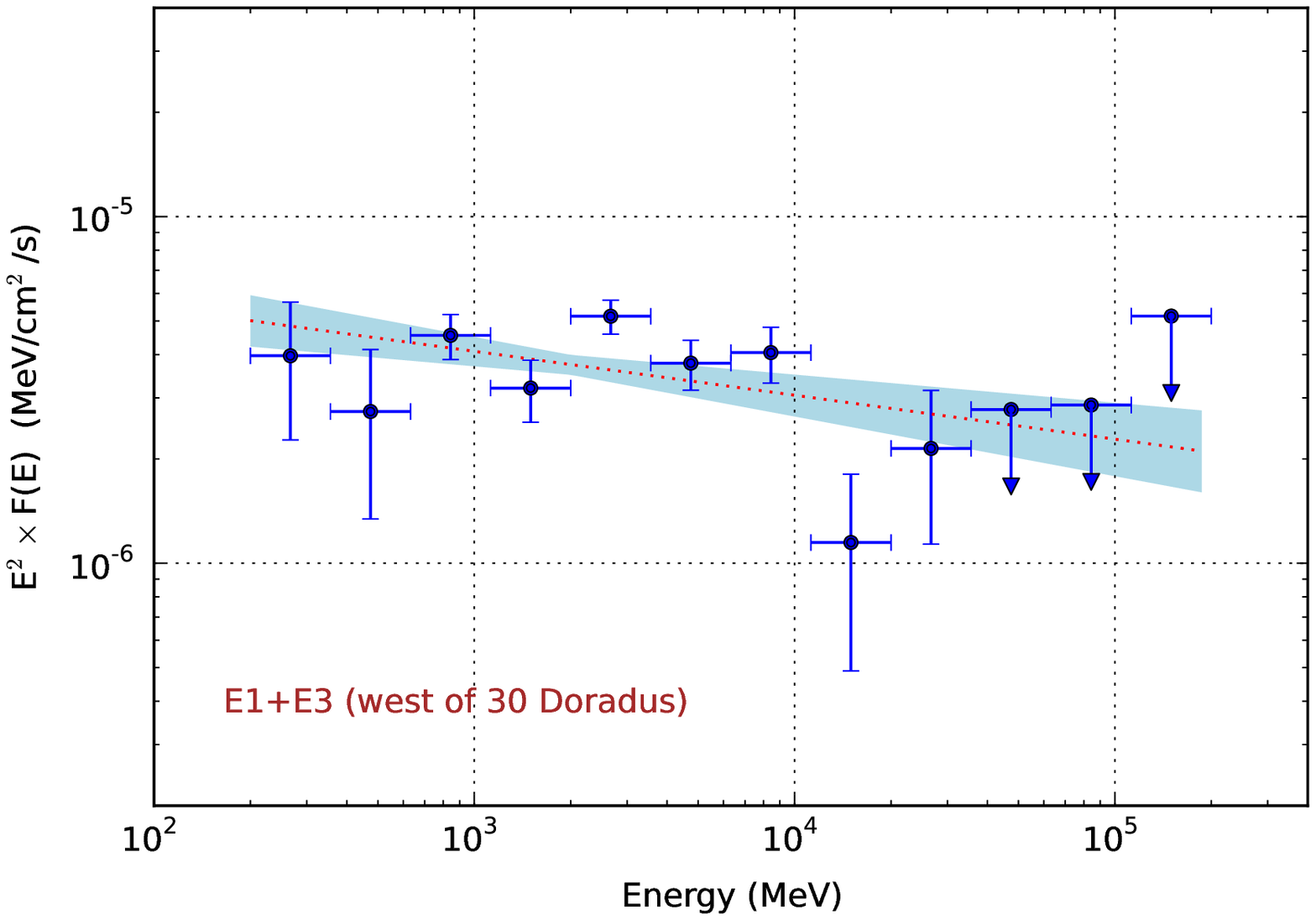}
\includegraphics[width= 7.8cm]{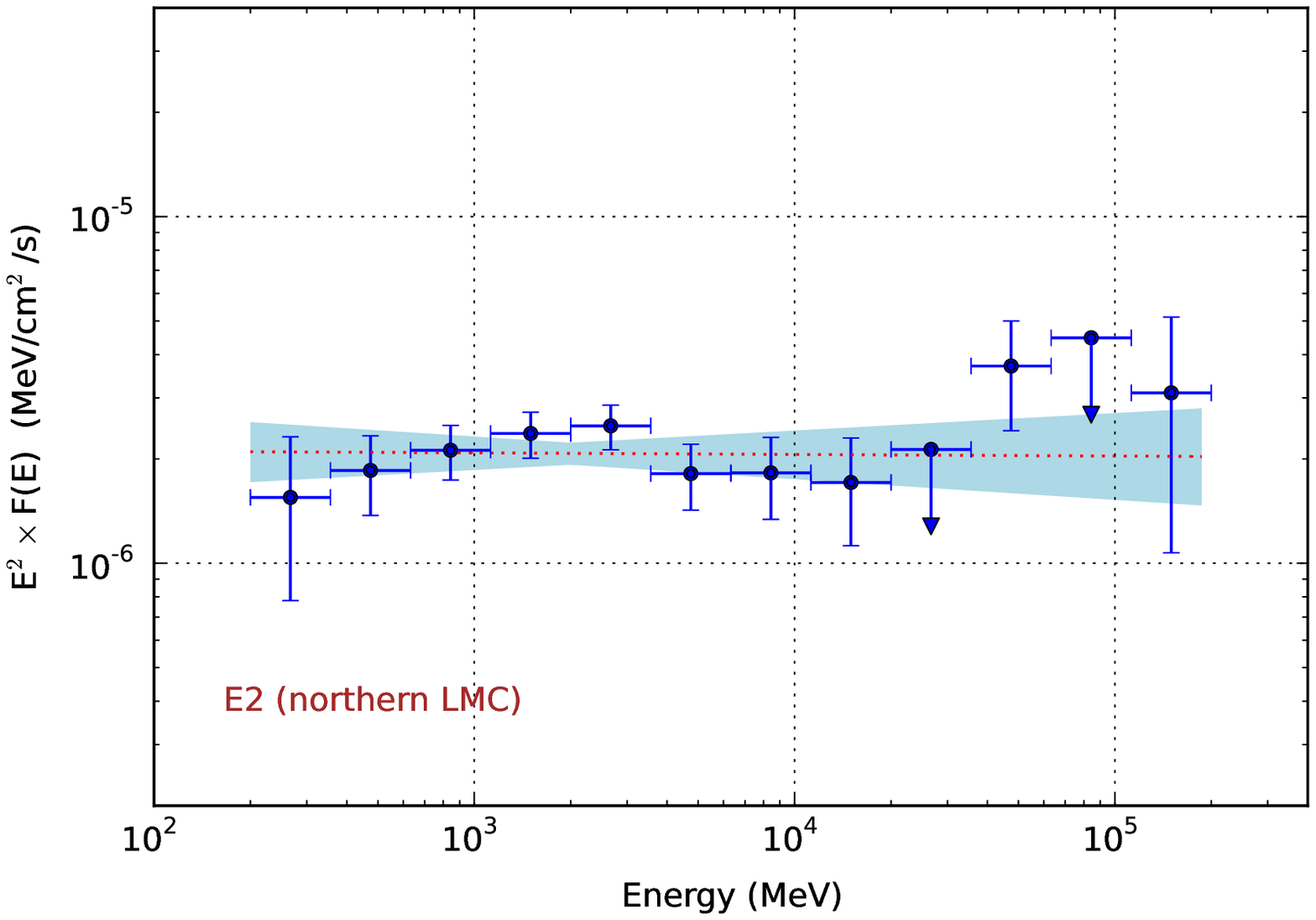}
\includegraphics[width= 7.8cm]{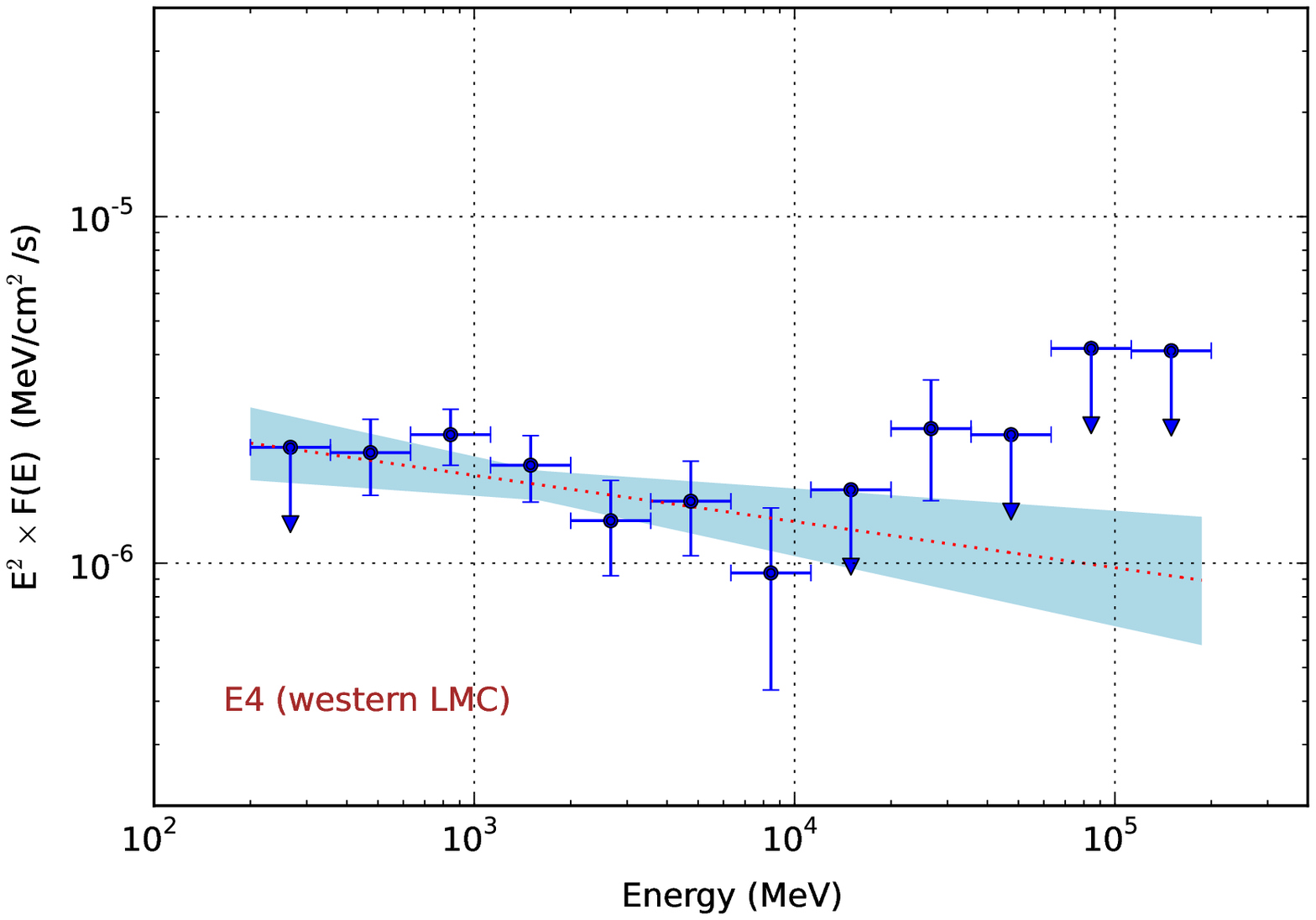}
\caption{Same as Fig. \ref{fig_spec_ptsrc} for the four extended emission components, from top to bottom E0, E1+E3, E2, and E4, as listed in Table \ref{tab_emimodel}. The spectrum of E0 has five bins per decade and a single upper limit above 20\,GeV computed for a fixed power-law photon index of 2.7; the spectra of other components have four bins per decade.}
\label{fig_spec_ext}
\end{center}
\end{figure}

The spectrum of the large-scale disk component labelled E0 shows some curvature (see Fig. \ref{fig_spec_ext}). We searched for the best model for it among a power law with exponential cutoff, a broken power law, a log-parabola function, and a tabulated function derived from the local gas emissivity spectrum (as a result of its interactions with CRs). Results are summarised in Table \ref{tab_compspec}. The best analytical model is the log-parabola. The broken power law is almost as good ($\log \mathcal L$ decreases by 1 despite 1 additional degree of freedom), while the power law with cutoff is disfavoured ($\log \mathcal L$ decreases by 3 for the same number of degrees of freedom). The significance of the curvature allowed by the log-parabola model compared to the power law is $7.4\sigma$. The tabulated function model slightly improves upon the log-parabola, with a likelihood increase by about 1 for 2 fewer degrees of freedom. At this stage, however, we avoided being too specific regarding the emission associated with component E0. We therefore used the best-fit log-parabola function, with spectral parameters listed in Table \ref{tab_extspec} and corresponding spectrum shown in Fig. \ref{fig_spec_ext}. An interpretation in terms of CRs interacting with the gas is presented below, among other possibilities, and the use of the tabulated function as a spectral model is discussed there (the similarity between the log-parabola and tabulated function models can be inferred from Fig. \ref{fig_emiss_spec}).

The spectra of components E1+E3, E2, and E4 can be described with simple power laws having relatively hard photon indices of $\sim$2.0--2.2 (see Fig. \ref{fig_spec_ext}). We evaluated whether these small-scale components can be distinguished from large-scale component E0 in terms of spectral properties. At low energies $<2$\,GeV, they have spectra consistent with that of E0 (a power law with photon index 1.9, as inferred from the broken power-law fit). At energies  $>10$\,GeV, only E2 and E4 are significantly detected (with TS $>20$), while E1+E3 and E0 are not\footnote{That components E2 and E4 could be detected above 10\,GeV at levels $\sim 3 \times 10^{-6}$\,MeV\,cm$^{-2}$\,s$^{-1}$ while component E0 is only constrained by an upper limit at a similar magnitude is due to their smaller size; this results in higher surface brightnesses, such that the smaller-scale components can be detected against the diffuse background at comparatively lower fluxes.}. In a binned likelihood fit over the entire energy range, replacing the power-law assumption for components E1+E3, E2, and E4 by the tabulated function model tested on component E0 yields the following results: the fit is hardly degraded for E4, with variations in $\log \mathcal L$ that are smaller than the number of degrees of freedom lost; in contrast, the fit for components E1+E3 and E2 degrades by about 10 and 20 in $\log \mathcal L$, respectively, for 1 degree of freedom lost. These results might seem to contradict the fact that E4 is significantly detected above 10 \,GeV, while E1+E3 is not. The spectra of these two components appear somewhat irregular in Fig. \ref{fig_spec_ext}, when compared for instance to those of point sources (for E4, this is more obvious when using a finer energy binning). A possible explanation is that these extended components encompass different sources with different spectra (as considered in Sect. \ref{res_ext_unres}), such that a single power law or the tabulated model are just first-order approximations.

The total energy flux from the extended emission components in our model amounts to $1.2 \times 10^{-4}$\,MeV\,cm$^{-2}$\,s$^{-1}$ over the 0.1--100\,GeV range. In Fig. \ref{fig_tot_spec}, the total spectrum of that emission is plotted against the spectra of the point sources for which we have an established (\object{PSR~J0540$-$6919}) or likely (\object{PSR~J0537$-$6910} and \object{N~132D}) association with an LMC object. To build this total spectrum, we added the best-fit spectral functions of all extended components together. Emission of extended nature clearly dominates the GeV emission from the LMC. It is interesting to note, however, that the summed emission from only two discrete sources (\object{PSR~J0540$-$6919} and the source coincident with \object{PSR~J0537$-$6910}) amounts to about one third of the extended emission flux at 100\,MeV, and to about 20\% of it at 10\,GeV, and that the total point-like emission has a similar spectrum to the total extended emission.

\subsection{Cosmic-ray population}
\label{res_ext_emiss}

The emissivity model introduced above to describe the extended emission from the LMC provides direct estimates of the emissivity distribution over the galaxy. Before addressing this quantitatively, however, we wish to emphasise that this is an assumption about the nature of the extended emission we observe: it predominantly results from hadronic interactions between CR nuclei and interstellar gas. The increase in likelihood it leads to, compared to the more phenomenological description of the analytic model, seemingly supports that interpretation, but this is not a formal proof that the latter is valid. There are other possible contributions to the extended emission, such as unresolved populations of discrete sources such as pulsars, or inverse-Compton emission from CR electrons. They were not tested against the data because their modelling is much more uncertain, but they are addressed below and in Sect. \ref{res_ext_unres}. For this reason, the emissivity estimates given below should be considered as upper limits.

\begin{figure}[!t]
\begin{center}
\includegraphics[width= 7.8cm]{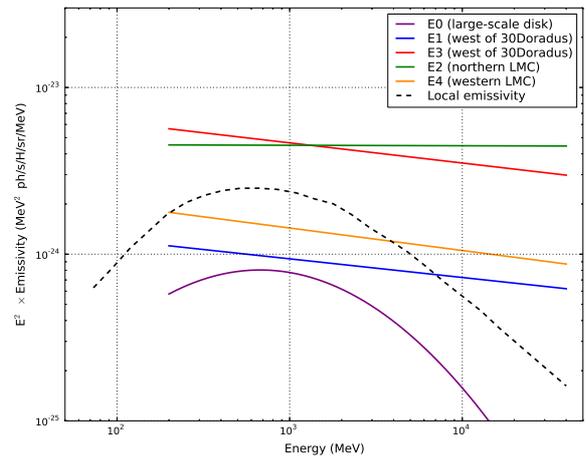}
\caption{Emissivity spectra of the five components of the emissivity model, obtained by dividing the fitted spectrum for that component by the average column density over the region, weighted by the 2D Gaussian emissivity profile. The dashed line is the local Galactic emissivity spectrum from \citet{Casandjian:2012}, given here for comparison.}
\label{fig_emiss_spec}
\end{center}
\end{figure}

For each extended emission component of the emissivity model, the gas emissivity spectrum is obtained by dividing the fitted spectrum for that component by the average column density over the region, weighted by the 2D Gaussian emissivity profile. The resulting emissivity spectra are plotted in Fig. \ref{fig_emiss_spec}. 
For comparison, we plotted the local Galactic emissivity spectrum determined by \citet{Casandjian:2012} from high-latitude {\em Fermi}-LAT observations in the $\sim$50\,MeV--40\,GeV range.

In terms of spectral shape, the emissivity spectra are not all consistent with that measured locally in the Milky Way. As mentioned in Sect. \ref{res_ext_spec}, component E4 has a spectrum consistent with the local emissivity, while E1+E3 and E2 significantly differ from it. Splitting E1+E3 into independent components E1 and E3 reveals that E1 is consistent with the local emissivity spectrum, while E3 is not. That the emissivity spectral shapes of E1 and E4 are consistent with the local one suggests that the spectral distribution of CRs is similar in these regions and in the local medium, at least up to CR energies of a few hundred GeV (responsible for emission at a few tens of GeV). In these conditions, the differences in the normalisation of the emissivity can be related directly to a difference in CR density, as discussed below.

In terms of normalisation, the emissivity components depart from the local value. We emphasise that emissivity components were modelled with 2D Gaussian spatial profiles, which implies a decrease of the emissivity away from the best-fit central position; in the following, the emissivities of the various components are given as peak values at the centre of the Gaussian profiles. The large-scale disk peak emissivity is 33\% of the local emissivity. As mentioned in Sect. \ref{model_emiss}, this component can be interpreted as resulting from CRs accumulated in the LMC over time scales that are long compared to the recurrence time of CR injection in the ISM (by supernovae). 
The difference in emissivity level means that the large-scale density of CRs is about one third of the local Galactic density in the inner regions of the LMC and decreases to about 10\% in the outskirts (by construction, because the size of the component was fixed). At 1\,GeV, the emissivities of components E1 and E4 are 40\% and 60\% of the local emissivity, respectively. Although this implies an underdensity in CRs with respect to the local medium, it means an enhancement in CR density with respect to the large-scale peak level in the LMC by a factor 3 for component E4. Components E2 and E3 have an emissivity that is twice the local value at 1\,GeV, or about 6 times the large-scale peak level in the LMC. Because of the significantly harder emissivity spectrum of the E2 component, the enhancement in CR density increases with energy, up to about 8 times the local value at 10\,GeV or 24 times the peak large-scale level in the LMC. The main source of uncertainty on these emissivity estimates is the gas content of the LMC (see Appendix A).

\newpage
\begin{figure}[H]
\begin{center}
\includegraphics[width= 7.2cm]{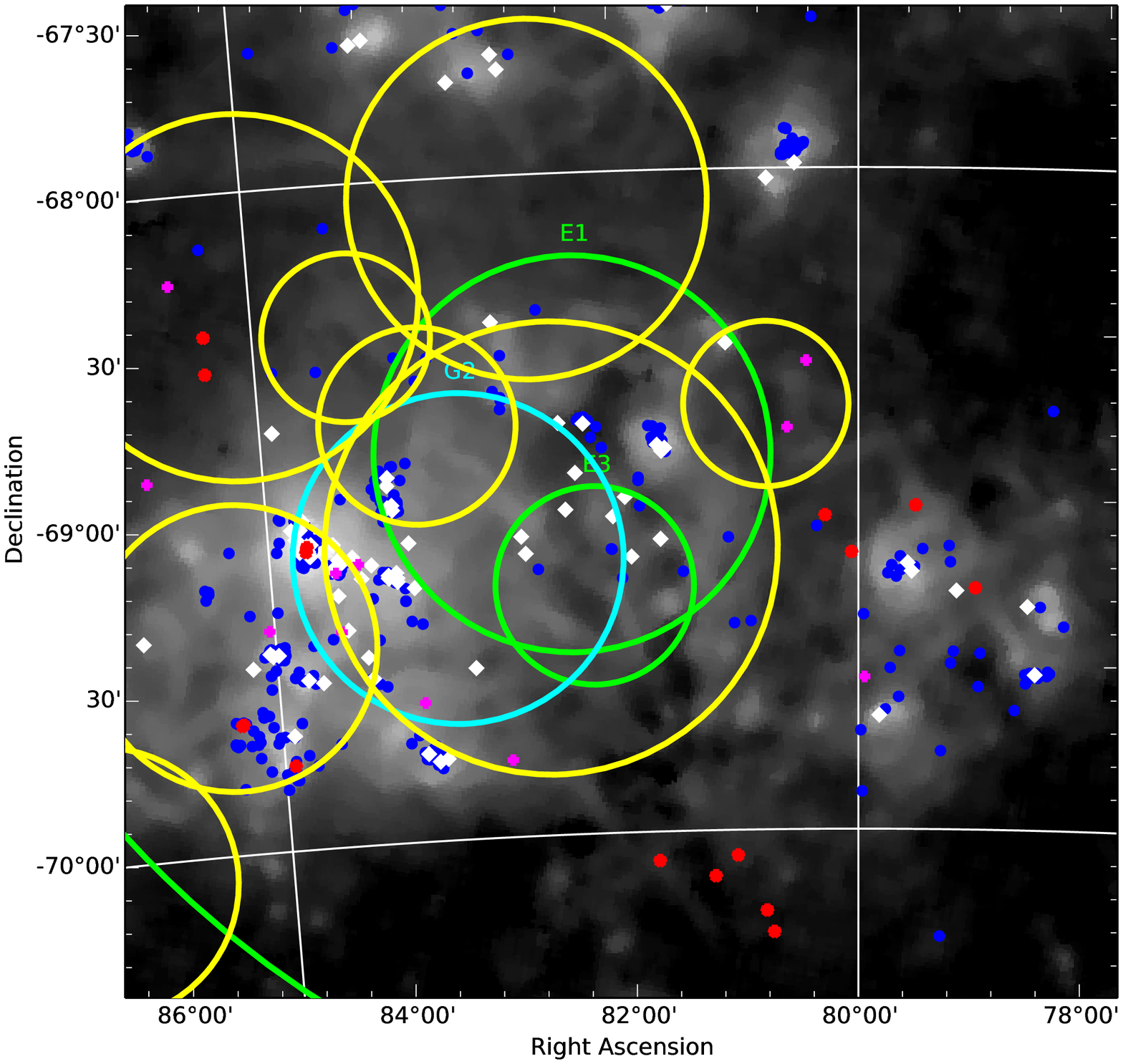}
\includegraphics[width= 7.2cm]{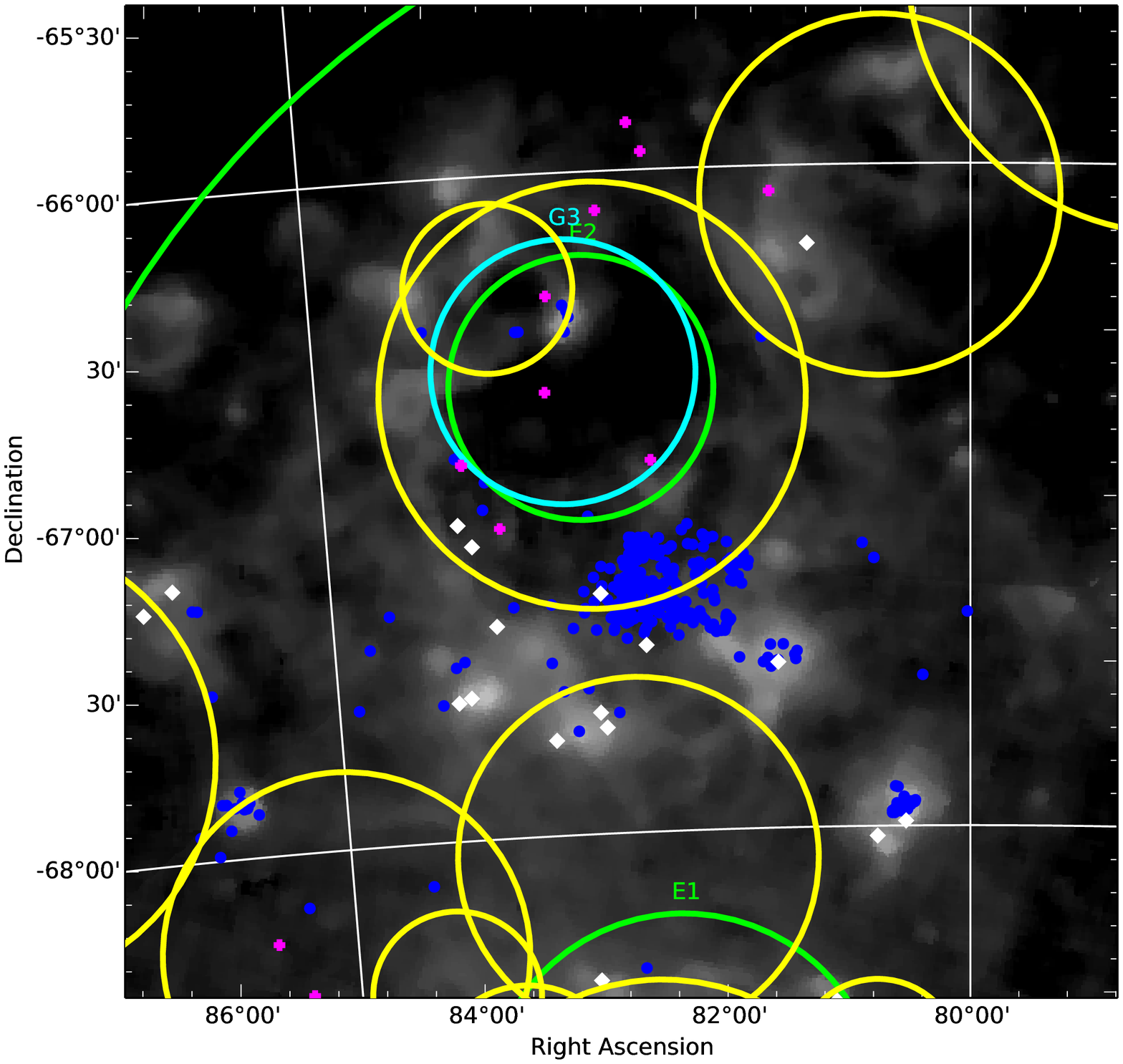}
\includegraphics[width= 7.2cm]{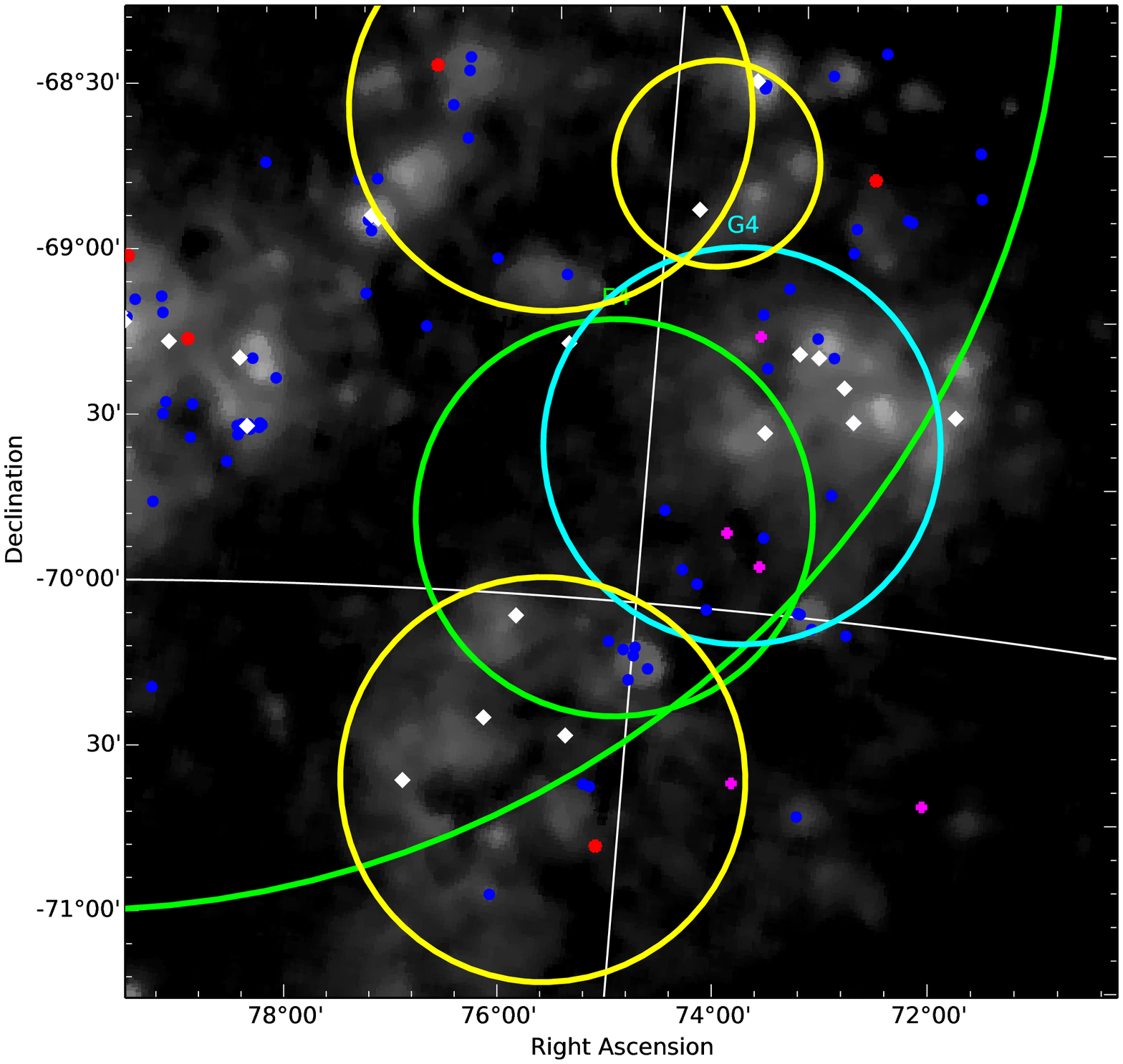}
\caption{Charts illustrating the layout of each region associated with small-scale extended \gray\ emission. From top to bottom: regions around emissivity components E1+E3, E2, and E4. The background is a SHASSA H$\alpha$ smoothed map on a log scale. The green circles are the 1-$\sigma$ extent of the Gaussian emissivity components of the emissivity model, the cyan circles are the 1-$\sigma$ extent of the Gaussian emission components of the analytic model. Overlaid are pulsars (magenta pluses), Wolf-Rayet stars (white diamonds), stars of spectral type B0-3 (blue dots), HMXBs (red dots), and supergiant shells (yellow circles).}
\label{fig_loca_ext}
\end{center}
\end{figure}
\newpage

\newpage
\newpage

\begin{figure}[!t]
\begin{center}
\includegraphics[width= 7.8cm]{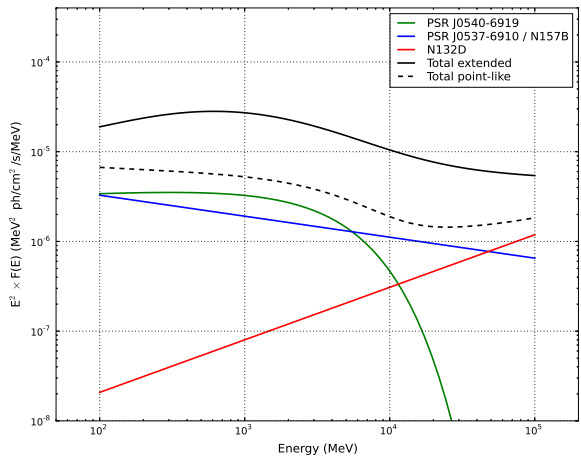}
\caption{Comparison of the total emission of extended nature with that of point sources for which an established or likely association with an LMC object exists (as indicated in the legend). Best-fit analytic spectra were used and summed over components E0 to E4 to produce the black curve for the diffuse contribution.}
\label{fig_tot_spec}
\end{center}
\end{figure}

Interestingly, components whose spectra are compatible with the local emissivity are those with lower normalisations, while components that are significantly harder are associated with higher normalisations. The high CR densities and harder emissivity spectra in components E2 and E3 suggest an accumulation of fresh CRs in these regions; these CRs would have been released relatively recently by a spatial and temporal concentration of sources, and they would not have had the time to diffuse away substantially, which preserves their hard injection spectrum (reflected in the emissivity spectrum). This situation, and the possible association of emission components with cavities and supergiant shells (see Sect. \ref{res_ext_loca}), is reminiscent of the Cygnus ``cocoon'' found in the Milky Way \citep{Ackermann:2011b}, although on much larger spatial scales. But it is difficult to firmly establish such a scenario, especially given the lack of clear spatial correlation with other objects (see Sect. \ref{res_ext_loca}). Overall, the CR population inferred above corresponds to $1.4 \times 10^{53}$\,erg in the form of $\sim$1--100\,GeV CRs in the disk, with an additional $0.9 \times 10^{53}$\,erg contained in the smaller-scale regions; given the supernova rate in the LMC, this implies a $\sim$1\,Myr residence time for CRs in the LMC disk (see Appendix B).

An alternative explanation for the relatively hard spectra of extended components is that inverse-Compton emission dominates the signal from these sources. In models of the steady-state CR-induced interstellar emission from star-forming galaxies \citep{Myself:2014}, the inverse-Compton spectrum is harder than the pion decay spectrum above a few 100\,MeV and up to a few 100\,GeV, and it can account for a significant fraction $\sim$20\% or more of the 0.1--100\,GeV flux. 
From the models of \citet{Myself:2014}, a small synthetic galaxy was found to be a possible representation of the LMC\footnote{The model is an atomic gas disk of 2\,kpc radius with a molecular core of 100\,pc and a density $n=20$\,H$_2$\,cm$^{-3}$.}. It yields a total interstellar differential flux of $2.5 \times 10^{-5}$\,MeV\,cm$^{-2}$\,s$^{-1}$ at 1\,GeV, consistent with the observed value (see Fig. \ref{fig_tot_spec}). The corresponding inverse-Compton emission spectrum was found to fit the {\em Fermi}-LAT spectra of components E2 and E4 as well as power laws, and even to slightly improve the fit for component E1+E3 (by about 2 in $\log \mathcal L$, for 1 fewer degree of freedom); in contrast, the spectrum of the large-scale E0 component is poorly represented by the inverse-Compton spectrum, with a degradation of 62 in in $\log \mathcal L$, for 2 fewer degrees of freedom. From the spectral point of view, the small-scale extended components are therefore consistent with an inverse-Compton origin of the emission. From the spatial point of view, however, the locations of these components are somewhat unexpected, at least for components E2 and E4. These are found in regions of the LMC that are relatively far away from strong sources of radiation, infrared or optical, and that are not known for particularly high strengths of the interstellar radiation field \citep{Bernard:2008}. In the absence of particularly abundant targets for the inverse-Compton scattering, emission from these regions would therefore require localised enhancements of the CR electron density, similar to the CR nuclei enhancement considered previously. We then face the same problem that these regions are not clearly correlated with particularly high concentrations of massive stars or SNRs or any other possible tracer of the past CR injection activity (see Fig. \ref{fig_loca_ext}). Independent of this consideration, such accumulations of high-energy electrons may be expected to be clearly detectable in radio synchrotron emission (provided the magnetic field is not anomalously low), but nothing particular is observed \citep{Mao:2012}. The case of component E1+E3 seems different. Being closer to the \object{30 Doradus} region and also near the so-called optical bar, the situation is more favourable regarding the availability of high-energy electrons and target photons. 

\subsection{Unresolved population of sources}
\label{res_ext_unres}

As suggested earlier, extended emission from the LMC may arise from an unresolved population of low-luminosity \gray\ sources instead of CRs interacting with the ISM. This may be especially true for emission from the small-sized regions (G2 to G4 in the analytic model , E1 to E4 in the emissivity model). The number of such objects does not need to be large. As an example, model component E2 for extended emission in the northern LMC should be replaced by at least five point-like sources with $TS$$\sim$25--50 to provide a similar fit likelihood (but the number of degrees of freedom involved in searching for the best-fit properties of these point-like sources is four times that needed for a single extended component). Components E1 to E4 have similar fluxes, so $\geq$20 objects would be needed to account for their cumulated emission.

Since the dominant class of identified GeV sources in the Milky Way is pulsars \citep{Acero:2015}, it is reasonable to assume that two dozen \gray\ pulsars with luminosities much lower than that of the exceptional \object{PSR~J0540$-$6919} could appear as extended emission from specific regions of the LMC. Interestingly, localisation maps in Fig. \ref{fig_loca_ext} already show a handful of known pulsars that coincide with the extended emission components: 11 out of 25 pulsars are found within the 1-$\sigma$ angular radius of regions G2 to G4. 
Many others may lie there and be currently undetected in radio and X-ray surveys \citep[an estimated number of potentially observable pulsars in the LMC is of the order of several $10^4$; see][]{Ridley:2013}.
Some of the extended emission components have hard spectra (for instance E2), however, with significant emission above 10\,GeV; this seems difficult to reconcile with the now established fact that the majority of middle-aged pulsars have spectra cutting off beyond a few GeV \citep{Abdo:2013}. On the other hand, pulsars tend to be associated with PWN, which are expected to have hard spectra at GeV energies, and two out of the four sources found in the LMC are detected beyond 10\,GeV.

An unresolved population of GeV sources may also include low surface-brightness SNRs.
The \gray\ luminosities and spectra of SNRs likely depend on many factors, including type of the progenitor, age of the remnant, and environmental conditions. In the Galaxy, the integrated emission from SNRs was put forward as a possible explanation for the observed harder 1--100\,GeV emission from the inner regions, compared to expectations based on large-scale interstellar radiation alone \citep{Volk:2013}. In the LMC, SNRs could therefore add their contribution to that of PWNe above 10\,GeV, where pulsar emission is expected to rapidly decrease, and thus account for the hard emission from some extended emission components. But SNRs and PWNe are relatively short-lived, with typical lifetimes of order $10^4$\,yr, and this is difficult to reconcile with the observation that some extended emission components seem to coincide with supergiant shells with ages of several Myr.

Overall, both unresolved populations of sources and CR-induced interstellar emission are possible explanations for at least some fraction of the extended \gray\ emission from the LMC. The two options are difficult to distinguish based on GeV observations alone, and considering observations at other energies will most likely be the key to clarifying the situation. For instance, future $\sim$TeV observations with the \textit{Cherenkov Telescope Array} may determine whether the fraction of the emission is indeed diffuse above 100\,GeV by detecting extended emission and/or by finding new discrete sources. At the other end of the electromagnetic spectrum, \textit{Planck} will soon provide a new view of the synchrotron emission from the LMC, thus revealing how the leptonic component of CRs is distributed, and future $\sim$GHz observations with SKA may refine the picture even more and increase the number of known pulsars in the LMC.

\section{Conclusion}
\label{conclu}

We have analysed {\em Fermi}-LAT observations of the LMC over about 73 months of operation. The analysis was made over the 0.2--100\,GeV range. After subtracting foreground and background signals, the \gray\ emission from within the boundaries of the LMC can be modelled as a combination of four point-like objects, a large-scale component about the size of the galaxy, and three
to four small-scale regions about 1--2\deg\ in angular extent. 

One of the point sources is unambiguously identified as pulsar \object{PSR~J0540$-$6919} through its characteristic pulsations; it is the first \gray\ pulsar detected outside the Milky Way and has the highest isotropic \gray\ luminosity. Another source is spatially coincident with plerion \object{N~157B} and its powerful pulsar \object{PSR~J0537$-$6910}; no pulsations from the latter are detected, and the relatively hard spectrum extending up to $>$50\,GeV suggests that the emission is not contributed by the pulsar alone, but may also be related to the TeV source detected with H.E.S.S.. A third point source is spatially coincident with the bright SNR \object{N~132D}, also detected at TeV energies by H.E.S.S.; this weak source is detected above 10\,GeV with a relatively hard spectrum that implies a broad peak in the spectral energy distribution at $\sim$100\,GeV. A fourth point source is currently unidentified and could be a background active galactic nucleus; it shows no sign of variability and its spectrum is soft with a photon index of $\sim$2.8.

The prominent \object{30 Doradus} massive star-forming region is quite bright in GeV \grays\ mainly because of the presence of \object{PSR~J0540$-$6919} and the source coincident with \object{PSR~J0537$-$6910}. In contrast, the second most active star-forming region in the LMC, N~11, does not show a remarkable enhancement of \gray\ emission on top of the large-scale extended emission. \object{SN1987A} and \object{30 Doradus C} are not detected.

In spite of resolving individual GeV sources in the LMC for the first time, emission of extended nature is still significantly detected and accounts for the majority of the LMC \gray\ flux. The nature of this extended emission remains unclear. In particular, the small-scale extended regions do not correlate with known objects or structures in the LMC, except perhaps having some relation to cavities and supergiant shells.

A model in which the extended emission is interpreted as interactions of a spatially inhomogeneous population of CRs with interstellar gas provides a satisfactory fit to the data. Under this assumption, the LMC would be filled with a large-scale population of CRs with a central peak density of one third the local Galactic density; in addition, the small-scale extended emission components would be regions where the density of CRs is enhanced by factors of 2 to 6 at least. The \gray\ emission from some of these regions features relatively hard spectra, which can be interpreted as resulting from a more energetic and possibly younger population of CRs compared to that filling the LMC on large scales. An alternative scenario for the origin of the small-scale extended emission is that of an unresolved population of sources. Two dozen or more weak point sources can mimic the extended nature of the emission. Currently unknown pulsars and their associated nebulae or supernova remnants would be obvious candidates. Observations at other wavelengths, in particular in radio and at very high energies, will most likely be the key to clarifying the situation.

\begin{acknowledgement}
The \textit{Fermi} LAT Collaboration acknowledges generous ongoing support from a number of agencies and institutes that have supported both the development and the operation of the LAT as well as scientific data analysis. These include the National Aeronautics and Space Administration and the Department of Energy in the United States, the Commissariat \`a l'Energie Atomique and the Centre National de la Recherche Scientifique / Institut National de Physique Nucl\'eaire et de Physique des Particules in France, the Agenzia Spaziale Italiana and the Istituto Nazionale di Fisica Nucleare in Italy, the Ministry of Education, Culture, Sports, Science and Technology (MEXT), High Energy Accelerator Research Organization (KEK) and Japan Aerospace Exploration Agency (JAXA) in Japan, and the K.~A.~Wallenberg Foundation, the Swedish Research Council and the Swedish National Space Board in Sweden. Additional support for science analysis during the operations phase is gratefully acknowledged from the Istituto Nazionale di Astrofisica in Italy and the Centre National d'\'Etudes Spatiales in France.
This research has made use of the SIMBAD database, operated at CDS, Strasbourg, France.
\end{acknowledgement}

\bibliographystyle{aa}
\bibliography{/Users/pierrickmartin/Documents/MyPapers/biblio/SMC,/Users/pierrickmartin/Documents/MyPapers/biblio/CosmicRaySources,/Users/pierrickmartin/Documents/MyPapers/biblio/CosmicRayTransport,/Users/pierrickmartin/Documents/MyPapers/biblio/DataAnalysis,/Users/pierrickmartin/Documents/MyPapers/biblio/Fermi,/Users/pierrickmartin/Documents/MyPapers/biblio/Books,/Users/pierrickmartin/Documents/MyPapers/biblio/Physics,/Users/pierrickmartin/Documents/MyPapers/biblio/StarFormingGalaxies,/Users/pierrickmartin/Documents/MyPapers/biblio/Starbursts,/Users/pierrickmartin/Documents/MyPapers/biblio/Pulsars,/Users/pierrickmartin/Documents/MyPapers/biblio/LMC,/Users/pierrickmartin/Documents/MyPapers/biblio/CosmicRayMeasurements}
\newpage

\begin{appendix}

\section{Uncertainties on gamma-ray emissivity estimates for the LMC}

In the estimates of CR relative densities in Sect. \ref{res_ext_emiss}, we neglected the difference in metallicity between the LMC and the local medium because the corrections due to species heavier than He are small \citep{Mori:2009}. A major uncertainty on the above estimates comes from the uncertainty on the gas content of the LMC.  According to the discussion in \citetalias{Abdo:2010d}, the total gas mass in the LMC is quite uncertain because of a possible substantial amount of dark gas that is not revealed by the usual radio line tracers, but correlates well with the atomic gas distribution according to infrared observations. Following \citetalias{Abdo:2010d}, we adopted a total gas mass of $(7.2 \pm 2.4) \times 10^8$\,\msol. The uncertainty on the gas content may vary from one region to another in the LMC, but for simplicity we assumed that all regions have a $\pm33$\% relative uncertainty of their gas content. This directly translates into a $\pm33$\% relative uncertainty of the inferred emissivities. The choice of the $X_{CO}$ factor is also formally a source of error, but it is minor compared to the uncertainty on the total amount of gas (especially since extended emission is found in regions relatively poor in molecular gas).

\section{Energetics and residence time of the cosmic-ray population in the LMC}

The emissivities derived in Sect. \ref{res_ext_emiss} can be used to evaluate the total energy of the CR population in the LMC. Emissivity in the $\sim$1\,GeV range results predominantly from the decay of neutral pions produced by $\sim$10\,GeV CR protons, and this range of CR energies accounts for most of the CR energy density in the local ISM. The relative normalisations of emissivities at 1\,GeV are therefore interpreted as proportional variations of the CR energy density in these regions with respect to the local value.

This estimated energy density is then integrated over volume to yield an estimate of a total energy in CRs. The 2D Gaussian emissivity profiles were integrated up to $2\sigma$, except for
the E0 component, for which $1\sigma$ already covers the entire galaxy. The physical scale radii were determined from the best-fit angular size $\sigma$ of the 2D Gaussian spatial profile, assuming a $d=50$\,kpc distance to the LMC and correcting for the $i=30\deg$ inclination. An LMC disk thickness of $h=400$\,pc was used in the calculation \citep[the scale height of the atomic gas was estimated at about 200\,pc; see][]{Kim:1999}. For the large-scale disk component, which has an emissivity three times lower than local, this gives
\begin{equation}
U_{\mathrm{CR,E0}} = 1.4 \times 10^{53} \left( \frac{U_{\mathrm{CR,local}}}{1\,\mathrm{eV}\,\mathrm{cm}^{-3}} \right) \left( \frac{h}{400\,\mathrm{pc}} \right) \, \mathrm{erg}
\label{eq_crnrj}
,\end{equation}
where we assumed a local energy density of $\sim$1--100\,GeV CRs of 1\,eV\,cm$^{-3}$. This corresponds to the cumulative production of CRs by about 1400 supernovae, assuming each explosion releases $10^{51}$\,erg of kinetic energy, of which about 10\% go into the acceleration of CRs \citep{Caprioli:2014}. The supernova rate in the LMC being about 0.2\,SN/century \citep{Hughes:2007}, these 1400 supernovae imply an accumulation of $\sim$1--100\,GeV CRs over a typical lifetime of 0.7\,Myr. This is much less than the estimated 10--15\,Myr mean residence time of $\sim$1--100\,GeV CR protons in the Milky Way disk \citep{Brunetti:2000}.

To first order, this factor $\sim$10--20 difference on the residence time of CRs in the Milky Way and LMC is consistent with CR density estimates. The large-scale peak density in the LMC was found to be 30\% of the local Galactic density, which means about 20\% of the peak Galactic density \citep[there is a Galactocentric gradient in CR density and the peak value is $\sim$1.5--2.0 times the value at the solar circle; see][]{Ackermann:2011}. There is therefore a factor $\sim5$ difference in large-scale CR density, and not 10--20 as might be suggested by the difference in residence time. The factor $\sim$2--4 difference can be accounted for from the fact that the star formation rate per unit volume is a factor of about 3 higher in the LMC than in the Galaxy\footnote{Using as star formation rates 2 and 0.2\wunit, and as disk radii 20 and 3.5\,kpc, for the Galaxy and LMC, respectively.}. It is also interesting to note that the $\sim$1\,Myr residence time for CRs in the LMC disk contrasts with the typical age of 5--10\,Myr of supergiant shells, if the latter are confining CRs in localised regions, as suggested above.

Performing the same calculation for the other regions in our emission model results in a total CR energy of $1.5 \times 10^{52}$\,erg, $3.2 \times 10^{52}$\,erg, $1.8 \times 10^{52}$\,erg, and $2.2 \times 10^{52}$\,erg for components E1, E2, E3, and E4, respectively (with the same dependencies on the local CR energy density and LMC disk thickness as in Eq. \ref{eq_crnrj}). In total, this makes about $2.3 \times 10^{53}$\,erg in $\sim$1--100\,GeV CRs, or the equivalent of the production of 2300 supernovae. This is probably a lower limit since we neglected the contribution of additional higher energy particles in these regions featuring a hard spectrum, and the amounts of CRs present at high altitude above the disk.

\end{appendix}

\end{document}